# The impact of economic policies on housing prices. Approximations and predictions in the UK, the US, France, and Switzerland from the 1980s to today

By Dr. Nicolas Houlié[*,1]

[1] ETH Zurich, Institute of Geophysics, Seismology and Geodynamics, Zurich, Switzerland

* Independent researcher

## Abstract

I show that house prices can be modeled using machine learning (kNN and tree-bagging) and a small dataset composed of macro-economic factors (MEF), including an inflation metric (CPI), US treasury rates (10-yr), Gross Domestic Product (GDP), and portfolio size of central banks (ECB, FED). This set of parameters covers all the parties involved in a transaction (buyer, seller, and financing facility) while ignoring the intrinsic properties of each asset and encompassing local (inflation) and liquidity issues that may impede each transaction composing a market. The model here takes the point of view of a real estate trader who is interested in both the financing and the price of the transaction. Machine Learning allows for the discrimination of two periods within the dataset. First and up to 2015, I show that, although the US treasury rates level is the most critical parameter to explain the change of house-price indices, other macro-economic factors (e.g., consumer price indices) are essential to include in the modeling because they highlight the degree of openness of an economy and the contribution of the economic context to price changes. Second, and for the period from 2015 to today, I show that to explain the most recent price evolution, it is necessary to include in the datasets of the European Central

Bank programs, which were designed to support the economy since the beginning of the 2010s. Indeed, unconventional policies of central banks may have allowed some institutional investors to arbitrage between real estate returns and other bond markets (sovereign and corporate). Finally, to assess the models' relative performances, I performed various sensitivity tests, which tend to constrain the possibilities of each approach for each need. I also show that some models can predict the evolution of prices over the next 4 quarters with uncertainties that outperform existing index uncertainties.

# 1. Introduction

Assessing the status of the housing market is key in assessing the stability of the financial sector and inferring the level of risk taken by the banks, insurance, private equity funds, and private persons. Therefore, it is crucial for banks, insurance, regulators, and governments to understand the real estate market. Market observation is mostly done through the monitoring of prices, checking vacancy rates, and discussing the right level of a given transaction ("What if the fair price of a flat in Zurich"). Observed values are used to infer model regulatory capital, financial exposures, and the prices of secondary market securities with mortgages as underlying. However, pricing real-estate assets is one of the most challenging tasks financial institutions must complete as price structures are not always transparent to them, and secondary investors are not always aware of the local economic context. Here, I take a chance to model data using Machine Learning (ML) methods to explore first which model variables are the most important to us and, secondly, to assess the performance of the more sophisticated models in explaining the prices observed. ML has already been used to solve similar problems in the past [1-8].

<u>Macro view and risk to the economy</u>

This study does not aim to solve the real-estate pricing problem definitely but rather links the observed house prices to an economic context that is wider than the local markets and single transactions. From a macroeconomic perspective, the status of the real estate market is considered an indicator of an economy's overall health and one of its drivers [9-11]. Market bubbles [12] and crashes [13] are known to impact the finances of large parts of the population through rent levels, prices of properties, or changes in available income [14, 15]. The real-estate market also sustains large portions of the economy through construction, building maintenance, legal and planning needs. When a housing price downturn happens (e.g., 1990s in Switzerland for flats; in 2008 globally), the level of economic activity of those actors (e.g., building companies, lumber industry) may be lowered, inducing a stagnation of the economy. Those economic issues could be transmitted to other markets Field (Nneji et al., 2013a; Yunus, 2018) and countries' Field (Nneji et al., 2015) if the economies are open enough. Those contagion may be enabled through lenders covering multiple housing market (insurances and banks), through currency crisis. Such contagion effects may indeed be encouraged through the support programs of central banks to regional economies, which are already known to have cross-border effects [16, 17]. Therefore, in each country, the economic and political spheres constantly contemplate and debate about the status of the real estate market.

Regulations and safety measures

The conclusions of these reflections result in sets of support measures and regulations. To protect buyers, minimum income levels are required while maximum duration and rates of loans are prescribed. For the financial system, real estate exposures require significant financing needs, and regulators request large amounts of capital to guarantee the stability of the banking and insurance sectors in case a systemic crisis hits the housing market. To estimate capital needs, credit rating models and risk management measures are established to ensure the capital level is sufficient at any time. In this context, the price of purchased assets and their evolution over time become important, even for secondary investors (e.g., mortgage-backed securities owners). Those measures nevertheless rely on the price estimation at all times of each asset. Exposures, Exposure at Default (EAD), Loss Given Default (LGD), and Probability of Default (PD) models are all impacted by a mispricing of assets (e.g., through Loan to Value, collateral value, recovery value) requiring the bank to rely on independent pricing of assets and additional risk measures (liquidity monitoring, margin calls, etc.).

A short literature review on real estate pricing

Published prices are often based on the asset's parameters (hedonic models: type of asset, living surface, number of rooms, etc.), but they also depend on financial and economic contexts, which may remain unclear to players. For instance, if mark-to-market prices are often thought to be wrong, it is because the rationale of a transaction is usually overridden by considerations based on different goals. The price of a single asset results from a mix of emotional ("location, location, location," prestige address, rare properties, etc.), cultural (renting vs. buying), and economic variables such as the structure of the pension investments, professional needs, and touristic locations [18-26]. As said above many of pricing models available today focus on explaining prices from asset features and local contexts: the GDP [27-30], the quality of the assets on the market, their distance to main economic centers, the volumes available, the occurrence of crime [31, 32], land quality [33], level of salaries [34], environmental standards, industrialization, building norms and standards, political context [27], the type of immigration [35], the COVID-19 pandemia [36-38], the level of inflation [39] and the quality of transportation networks which enable workers to reach their job [40]. Those models, called hedonic (see Fahrländer [41], Wuest and Partners [42], Bourassa, et al. [43] for Switzerland), certainly allow the ranking of assets within a local context but tend to neglect other effects, such as the impact on personal taxes and macroeconomic conditions, and lead to the mispricing of many assets [44-46]. Another root of mispricing is also the consequence of competition of multiple pricing on the same kind of asset: retail transactions also compete with more significant projects (multiple-family buildings or multi-building) carried out by insurance companies and banks, designed to generate cash flow while securing cash deposits from inflation fluctuations. Pricing may be influenced by the demand for buy-to-let investments of private investors. Older generations are making an arbitrage between real estate, collective investment, and anticipated donations to the youngest family members. From their perspective, a variety of mid-term investments are limited, so the purchase of a real estate asset may be prioritized over the profitability of the investment. At last, markets may be disrupted by the drafting of regulations made in the contexts of political and fiscal reforms, which are not always aligned with economic cycles. Mispricing has strong consequences and cannot be ignored as it is carried only by buyers. A too-high value would lead to a higher risk of default and potentially a charge in capital too high for banks. An underestimated price may not allow the transaction to happen and lead to markets being gripped.

Can ML bring something if you adopt a holistic perspective?

In this study, I take a step back from pricing complexity and adopt a holistic view of the problem, focusing on the fact that a market is made of the sum of realized transactions. After all, regardless of the level of financing rates, some contexts are not favorable to real estate investment (e.g., low liquidity of potential buyers). For instance, if investors worry about their economic future or the level of prices, they may not be inclined to invest their savings into securing housing. Similarly, investors unwilling to lose too much may refuse to finalize the transaction. Such a context was observed in France in 2023 when commercial banks did not immediately charge the cost of rates rise to their customers in an attempt to preserve the market. Nevertheless, this commercial measure, which aimed at preserving the bank's businesses from rate fluctuations, did not impede the building companies from being impacted by a substantial slowdown of transaction volumes. In short, whatever the market and the type of asset, each transaction needs to be completed by making available three parts: 1/ an asset (or a seller), 2/ a buyer, and 3/ a secured financing facility. Those three parties are convenient as they directly represent the transaction from a front desk perspective. Those parties carry all liquidity, financial constraints (willingness to sell, possibility to buy), economic confidence (financing rates and willingness to buy), and arbitrage (profitability of the transaction, investment into equities rather than real estate) that characterize such a transaction. I test here the assumption that the financing constraints strongly determine the price rather than the specifics of the asset transferred. With the help of ML, I aim at identifying which parameters are the most important to explain prices and and potentially to predict those.

## 2. Materials and Methods

For modeling the data, I used the caret R package [47, 48] and tested my approach in 15 countries (Table 6). I have tested both kNN and Tree-bagging strategies in order to detect the presence of overfitting in the model solutions. This paper is linearly organized by quantifying the impact quality of including new variable(s) to the previous model. But first, and based on a wide choice of data available at SNB, I select which indicators are helping the most to explain the evolution of flats in Switzerland. Second, and from the parameter selection, I have built an initial model based on inflation, financing rates, and GDP to explain the price evolution in 15 countries using the data already archived by the FRED database. I then test the inclusion of central bank rates and ECB portfolio size in order to test whether the addition of a parameter significantly improves the data fit. I have tried to model the house prices in nominal and percentage forms to test whether using the nominal value improves the model performance. The complete list of models is available in Table 1.

I have used three different sources for the data used in this study. First, I used the Swiss National Bank (SNB) dataset, which helped me to identify which parameters were critical to include. The dataset available at SNB is extensive and detailed, but I have failed to find the exact same data for other countries. For instance, price indicators for each canton and each asset type (flat vs single-family house) cannot be found for all countries. As I wanted to expand my study to a large number of countries, I used the data available at the Federal Reserve Bank of Saint Louis ("FRED") for economic and financial data and at the Federal Reserve Bank of Dallas House Price Index database [49].

Regarding the data preparation, the pre-processing was kept to a minimum. For all models, end-of-quarter figures were used for all data. When missing data were observed (the treasury rates data were not complete at the daily level), gaps were filled using linear interpolation when end-of-quarter values were missing. The original data and the corrected data are shown in the supplementary materials for treasury rates. No modification was applied to house prices when nominal values were used. I show the for Consumer Price Indices (CPI) were utilized in their annualized form. The same data treatment has been used for all the countries considered.

|   | Model name | GDP | CPI (rates) | Treasury 10y | Unemployment | Rent | Central bank | ECB | HPI | Figure |
|---|---|---|---|---|---|---|---|---|---|---|
| 1 | 3-parameter | Y | Y | Y, rates | N | N | N | N | Nominal | Figure 4 |
| 2 | 3-parameter 1yr | Y, rates | Y, rates | Y, rates | N | N | N | N | Rates | Figure 10 |
| 3 | IR models | Y, rates | Y, rates | Y, rates | N | N | Y, US rates | N | Nominal | Figure 5 |
| 4 | Local Central bank | *Y, rates* | *Y, rates* | *Y, rates* | *N* | *N* | *Y, local rates* | *N* | *Nominal* | *Figure 6* |
| 5 | ECB | Y, rates | Y, rates | Y, rates | N | N | N | Y, nominal | Nominal | Figure 7 |
| 6 | ECB/FED | Y, rates | Y, rates | Y, rates | N | N | Y, plus FED rates | Y | Nominal | Figure 8 |
| 7 | ECB 1yr | Y, rates | Y, rates | Y, rates | N | N | N | Y | Rates | Figure 11 |
| 8 | LOCAL | Y, rates | Y, rates | Y, rates | Y | N | N | N | Nominal | Figure 13 |
| 9 | LOCAL 1yr | Y, rates | Y, rates | Y, rates | Y | N | N | N | Rates | Figure 6 |
| 10 | Rents | Y, rates | Y,rates | Y, rates | N | Y | N | N | Nominal | Figure 9 |
| 11 | Rent 1yr | Y, rates | Y, rates | Y, rates | N | Y | N | N | Rates | Figure 9 |
| 12 | Permutations | Y, rates | Y, rates | Y, rates | N | N | N | N | Nominal | Figure 12 |

Table 1: Characteristic of the data used for each model group. HPI rates are computed over 12 quarters[1].

---

[1] 
$$\Delta HPI_i = \frac{HPI_i - HPI_{i-12}}{HPI_i}$$

## 2.1. The Swiss National Bank (SNB) data portal dataset

Many economic data for Switzerland are available at the Swiss National Bank (SNB) data portal (www.snb.ch). Real estate prices are available for various types of assets (flats, single-family houses, etc.) in multiple areas. Other data potentially valuable for this study relate to the macroeconomic trends of the Swiss economy (GDP, GNI, subsidies, wages, etc.). The published real-estate prices show the dispersion of estimates for each type of asset (flats, single-family homes, buildings) across the pricing tools (Figure 1). Values for some assets and some years show a dispersion of +/- 5% for homes and +/-15% for flats. For Wuest and Partner, the spread between "ask" and realized prices can reach 30% for some market segments. Gaps between price time series may reflect differences in the size of datasets, segment of the market, or regional location of the assets observed. Those differences will be discussed later in the light of fit performances and predicted price accuracy in the result section. To consistently include more countries in this study, I searched for data that could be found in other databases, such as INSEE in France. Unfortunately, data coverage using national agency datasets cannot provide enough consistency across the countries (some series are too short to be exploited, and others do not have a sampling rate similar to other countries).

a)

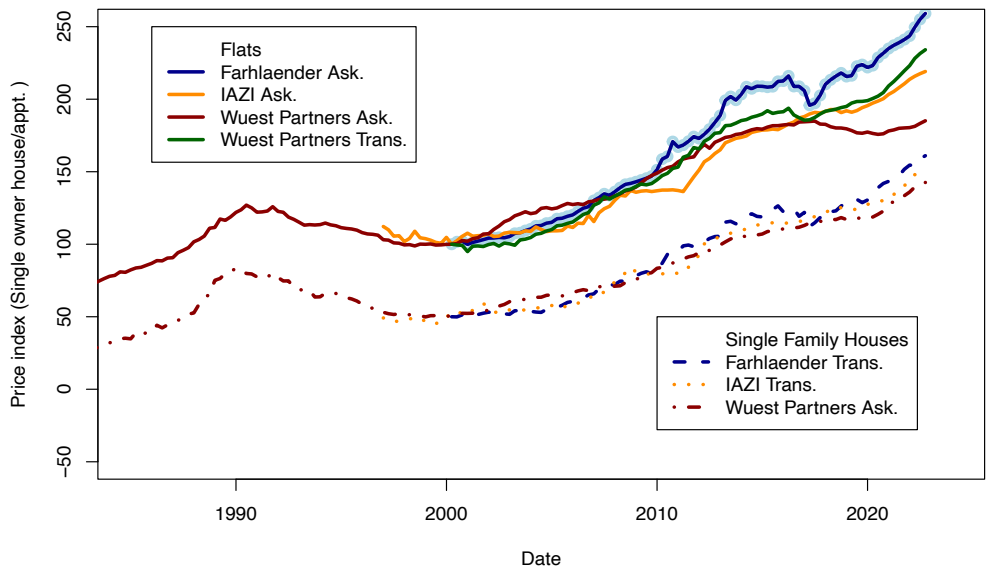

b)

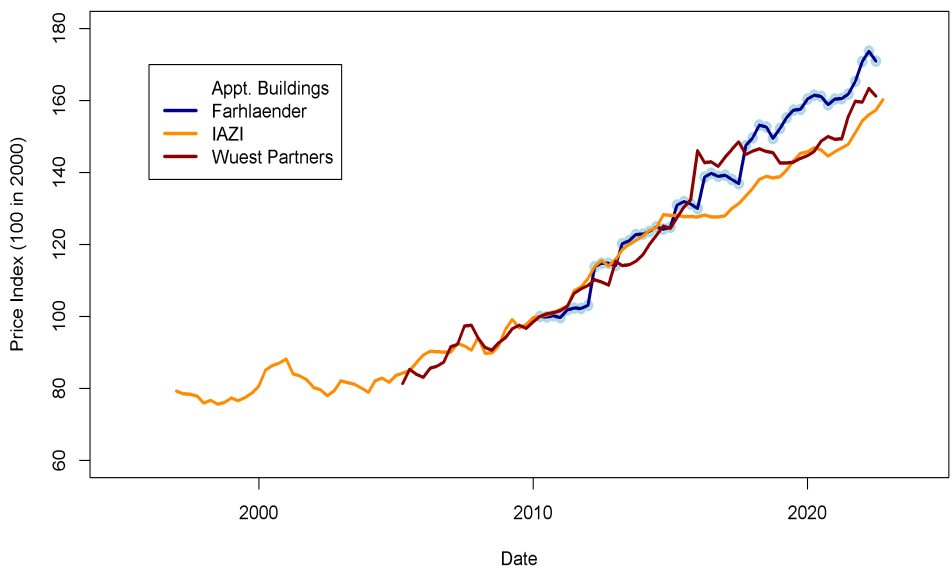

*Figure 1: Real-estate prices as published on the SNB data portal. a) Ask and Transaction prices for flats and single family houses in Switzerland. Prices for single-family houses have been shifted by minus 50 points for visibility. It is visible that indices are less scattered for house prices (+/- 5%) than for flats (+/- 15%). Those differences may be explained by various reasons such as geographic sampling, range of asset quality, etc. b) Transaction prices for apartment buildings in Switzerland in different indices. The standard deviation of index prices is close to +/-5%. Data sources: IAZI, Wuest Partners and Fahrlender.*

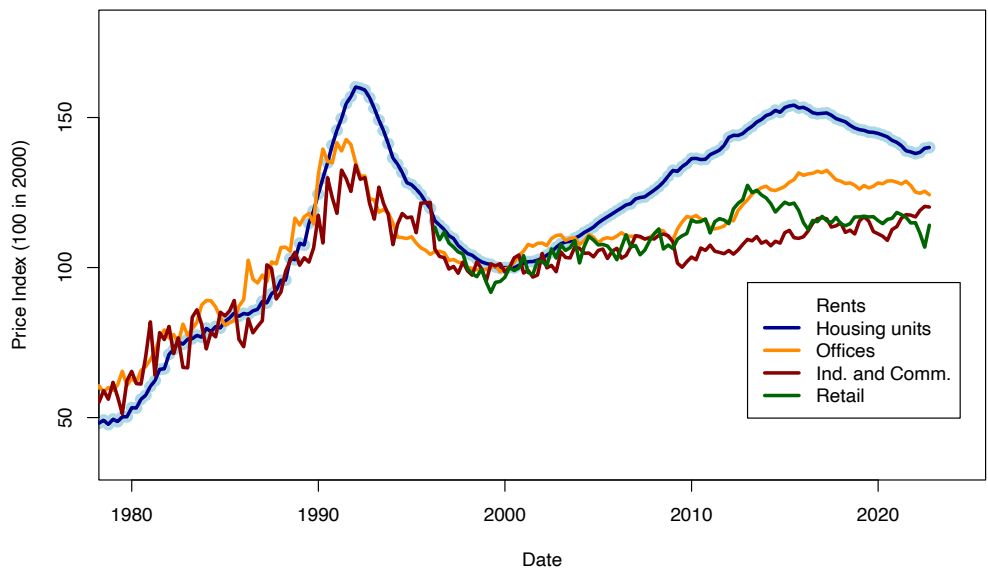

*Figure 2: Rent level series for various non-occupied assets. Data source: Swiss National Bank. The final model of this study will include the rent level for housing units (flats and single-family house). The rent model evolution follows well the evolution of prices at least until 2015. Further analysis may be necessary to understand why rents were decreasing in the years 2015-2020 while buy prices kept increasing.*

## 2.2. The FRED dataset

The Federal Reserve Bank of Saint Louis archives macroeconomic data for many countries in the same format and distributes it freely through its website. From the first analysis of the data for Switzerland, I chose the following parameters: Consumer Prices Index (CPI) values, US Treasury Rates (10-year maturity), and Gross Domestic Product (GDP) for 15 countries. To those parameters, I added the interest rates of the central banks of each country and the FED/ECB total asset portfolio size, which are known to impact the level of economic activity and house prices [50, 51].

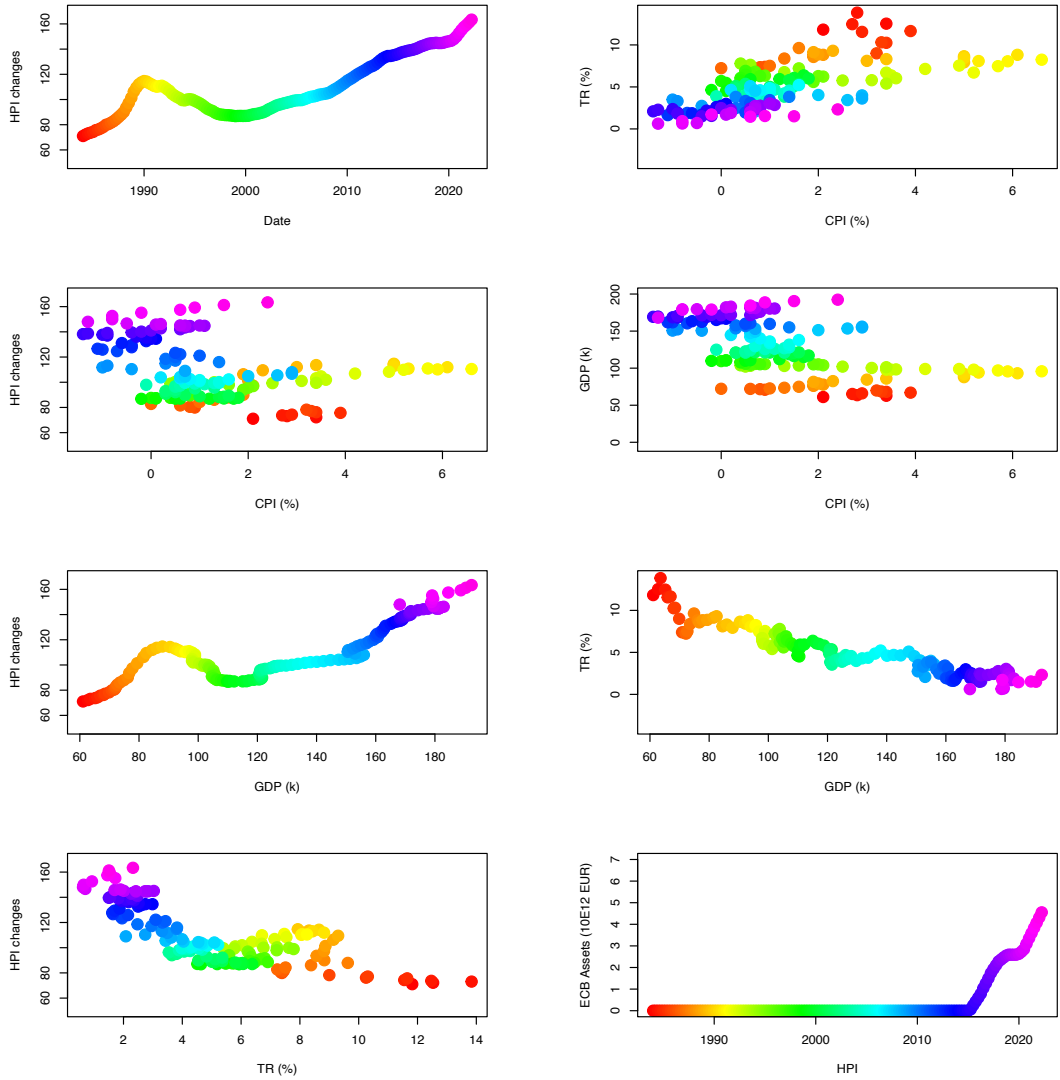

*Figure 3: Evolution of economic variables in Switzerland. a) HPI expressed as other parameters. The color code consistently symbolizes the date of each point across the panels (see first panel to obtain the correspondence between dates and color code). Inflation seem to be correlated to prices only over the long-term. From the analysis of the GDP series, prices seem to evolve independently of it. At last US treasury rates if they are correlated to prices, seem being to vary by many percent points without impacting the prices. Sources are the FRED database and HPI dataset.*

## 2.3. The International House Price Database of the Federal Reserve Bank of Dallas

At last, I opted to use the house prices published by the Dallas Federal Reserve [49]. The house price index (HPI) data set covers many countries (see list of countries included in this study; Table 6). I chose to use the HPI and not its inflation-adjusted version (RHPI) as I preferred using the inflation data (CPI) as an input parameter in the raw form. It is possible that the baskets of goods used to compute CPI differ from one country to another. It is not an issue here, as each CPI dataset is for house prices in the same country. Doing this allows for a better match between the local perception of prices and the confidence of buyer/sellers into entering a transaction. Regarding the choice of countries, I focus on the European countries and some countries of interest (USA, Australia, Canada) because their market is sizable and/or because they are known to use variable interest for part of the duration of the loan. By doing so, I hope to discover similarities between European countries and between countries with strong economic links (e.g. France-Germany; USA-United Kingdom; United Kingdom-Australia; Finland-Sweden-Norway). I show in (Figure 3) the evolution of some of the indicators used against the data published by the Federal Reserve Bank of Dallas House Price Indicator (HPI) for Switzerland.

# 3. Results

## 3.1. Two first models to map important variables for Switzerland

In this section, I complete two models that aim at finding the most important variable and quantifying the contribution of new data to the model improvement. I first use a dataset of 49 parameters available at the SNB. The complete list of variables is available in the appendix.

The data included in this model are unmodified from their source. HPI values are used in their nominal form with a reference value of 100 at Q2-2002.

R-caret allows for quantifying the relative importance of the parameters contributing to explaining the signal. I list the most essential 20 parameters with their categories in Table 2. Most of the parameters listed are either related to inflation or income. Interestingly, employment rates and short-term financing rates are playing a less important role than expected. For Switzerland specifically, these first findings are explained by the fact that only households with yearly income larger than 1/5 of the total loan are given access to financing (this is a thumb rule valid to investment purchased with a maximum of 30% collateral provided.

Overall, data shown in Table 2 are reassuring, and this model confirms that because the affordability criteria fixed by financial authorities are strictly enforced, the level of unemployment rates has no impact on prices. Some observations can be additionally made:

1. Financing rates (e.g. LIBOR 3M) are playing a role but less pronounced than expected. Most actors agree that the financing rate is the most critical parameter in explaining the evolution of house prices. In the view of many, low financing rates are allowing more transactions to be realized, driving the demand upwards. This is undoubtedly true for most buyers, but the transaction's performance also plays a role for investors involved in large volumes. For buy-to-rent schemes, real-estate investments become more profitable when rates are low. Indeed, as the profitability of a real-estate transaction is estimated at 3 to 5%, the arbitrage between a real-estate transaction and another investment that is equally cash-intensive (e.g. public or private debt) must be considered in the light of the inflation context for the period covered by the transaction (2,5 or 10 yrs).

2. Secondly, this first model shows that inflation plays a role. Inflation can have multiple effects on the health of a portfolio. Those effects depend on the real-estate segment:

- For private investors, inflation is effective at reducing the interest rate charge. For those sensitive to downward income fluctuations, a change in inflation may become a concern. For those with a fixed interest rate or those who use real estate investments to protect their savings, an inflation increase could be beneficial, especially if real estate prices follow inflation.

- For commercial and buy-to-let investors, inflation has primarily adverse effects. Indeed, those investors may have inflation-linked and floating loans. If the business plan done at inception was not conservative enough, inflation may impact the investor's liquidity through an increase in maintenance and operation costs (e.g., heating).

3. At last, I contemplate the impact of GDP on house price structure. In reality, GDP is not correlated with price evolution: the GDP never decreased in Switzerland while the HPI fluctuated. Nevertheless, we include GDP into each national dataset as a local parameter, weighting the overall perception of the regional economic conditions against the other countries. For instance, commercial investors may examine the GDP dynamics before choosing the location of an office and commercial space investment. However, it is unlikely that private investors will weigh their purchase decisions in light of GDP forecasts.

As a second step and to increase the number of quarters available in the dataset, I have reduced the set of parameters. Model 2 shows similar performance to model 1, although the number of data points is much smaller. From the comparison of the standard deviation of residual values, model 1 and 2 have similar performances (Figure 4).

As expected, these first two models show the power of the k-nearest neighbors machine learning approach in discriminating the importance of various input parameters to explain output data.

|   | Parameter Name (SNB denomination) | Category | Relative Importance |
|---|---|---|---|
| 1 | Compensation of employees paid to the rest of the world | Income | 100.00 |
| 2 | Type of goods - Goods - Durables | CPI | 99.55 |
| 3 | Type of goods - Services - Total | CPI | 99.49 |
| 4 | Compensation of employees | Income | 99.24 |
| 5 | Type of goods - Services - Private | CPI | 99.02 |
| 6 | Origin of goods - Domestic | CPI | 98.69 |
| 7 | Gross domestic product | Income | 97.09 |
| 8 | Gross national income GNI | Income | 96.95 |
| 9 | Subsidies | State | 96.78 |
| 10 | Consumption of fixed capital | Investment | 95.95 |
| 11 | Taxes on production and imports | State | 95.34 |
| 12 | Type of goods - Services - Public | Inflation | 91.33 |
| 13 | United Kingdom - GBP - GBP LIBOR - 3-month | Markets | 91.32 |
| 14 | Labour force | Demography | 88.95 |
| 15 | Compensation of employees received from the rest of the world | Income | 88.37 |
| 16 | Switzerland - CHF - Call money rate Tomorrow next - 1 day | Markets | 86.84 |
| 17 | CPI - Total index | Inflation | 84.31 |
| 18 | Net operating surplus | Production | 75.30 |
| 19 | Type of goods - Goods - Non-durables | Inflation | 73.91 |
| 20 | Property income paid to the rest of the world | Income | 72.37 |

*Table 2: The first 20 most contributing variables selected by the kNN algorithm from the complete dataset of 49 parameters for model 1. Those parameters in their original form (nominal values and percentages) are needed to explain apartment prices in Switzerland.*

However, inferring other datasets could have explained the data equally well is still impossible. We can, however, conclude from these two models that 1) the number of parameters to explain prices is not necessarily high, and 2) that treasury rates cannot be used as the only financial input parameters (i.e. otherwise it would not have been possible to finance real-estate in the 1970's).

For each model, and to rank them by their quality, I have computed various statistical figures such as Root Mean Square (RMS)[2], Standard deviation of residuals[3] (i.e. observed-modelled values), Mean Absolute Error (MAE)[4] and Mean absolute percentage error (MAPE)[5]. Regarding the MAPE values, I use the thresholds proposed by Lewis [52][6]. Those functions are part of the MLmetrics R package. The goal here is not to provide definitive answers to the price construction, but to explore which input parameters are significant enough to contribute explaining the data. For each model performance testing I run the models 600 times and build statistics from the fit of each model. I consider models are performing well when mean (RMS)< 15 and mean (MAPE) < 0.20.

| Model | M_COR | S_COR | M_RMS | S_RMS | SD | M_MAE | M_MAPE | M_CHIp | M_CHIs |
|---|---|---|---|---|---|---|---|---|---|
| Model 1 / 49v | 0.994 | 0.001 | 3.307 | 0.201 | 3.320 | 2.649 | 0.020 | 0.157 | 0.213 |
| Model 2/ "lightdata" | 0.974 | 0.001 | 0.464 | 0.009 | 0.466 | 0.373 | 0.110 | 0.095 | 0.702 |

*Table 3: Performance of the first two models based on Swiss National Bank dataset. The model ("lightdata"), which uses fewer input parameters but spans a more extended period, is more accurate (RMS, SD improve while COR and MAPE stay satisfactory). I consider both models are performing well (RMS< 15; MAPE< 0.2; COR>0.90). They are highlighted in bold font for this reason.*

---

[2] Root Mean Square (RMS): $RMS = \sqrt{\frac{\Sigma(x_i - obs.)^2}{N}}$

[3] Standard Deviation (SD): $\sigma = \sqrt{\frac{\Sigma(x_i - \mu)^2}{N}}$

[4] Mean Absolute Error (function MAE from R: MLmetrics package): $MAE = \frac{1}{n}\sum_{t=1}^{n}|\hat{y}_t - y_t|$.

[5] Mean Absolute Percentage Error (function MAPE from R: MLmetrics package; please note the x100 factor missing from the standard definition): $MAPE = \frac{1}{n}\sum_{t=1}^{n}\frac{|\hat{y}_t - y_t|}{y_t}$

[6] <0.1 Highly accurate forecasting, 0.1-0.2 Good forecasting, 0.2-0.5 Reasonable forecasting, and > 0.5 Inaccurate forecasting.

|   | Parameter Name (SNB denomination) | Category | Relative Importance |
|---|---|---|---|
| 1 | Compensation of employees | Income | 100 |
| 2 | GNI | Income | 96 |
| 3 | Inflation total index (%) | Inflation | 62 |
| 4 | US LIBOR 3m (%) | Rates | 24 |
| 5 | SNB Core inflation (%) | Rates | 0 |

*Table 4: List of parameters chosen for model 2. Here, as some parameters includes some collinearity with prices (positive for Compensation of employees and GNI; negative for US LIBOR 3m) the performance of this model should be considered with caution. However, this model presents the advantage to use raw data. This model also presents the advantage to link some demographics data such as immigration numbers (more workers may lead to a decrease of salary levels).*

| Model parameters | |
|---|---|
| tr control | method="repeatedcv", repeats = 3 |
| empty data | not allowed |

*Table 5: Training parameters used for all models in this study.*

### 3.2. Open Economies ?

It is well known that economic contagion may happen between countries and regions of the global economy. Contagions are made possible when regional and international institutions invested in multiple asset classes suddenly meet an unforeseen liquidity need or when the source of financing is unique for many markets. One obvious scenario would describe the difficulties of a commercial real estate investor who may be impacted by a sector recession (an increase of vacancies in rented assets), default of an investor (unfinished buildings), and/or liquidity difficulties of a lender forcing him to realize transactions at a loss (fire sale to reduce exposures) to increase capital ratios.

To test the connection(s) between real-estate markets, I computed the correlation coefficients between house price series of various countries (Table 6). Some series are correlated, others less. For instance, Italy, Spain, Russia, and Hungary are the countries least correlated with others. Japan's price history is anti-correlated with that of most countries. The overall correlation between different markets may be explained by a common source of financing, debt origin, and securitization, which makes the market more connected. Local conditions (demography, volumes available, construction rate, etc.) may play a more prominent role for countries that do not show a correlation. Countries publishing data only once a year (Japan and Russia) or only recently (Hungary) are excluded from further analysis as too few data are available. One can consider, however, that Japan, because of its specific context (demography, domiciliation of its debt, and status of its stock market), may stand out rightly. I build on these first observations and model independently house price time series for various countries by testing the impact of adding new variables on the quality of the data fit.

| | | Nobs | USA | FR | GER | SP | IT | FIN | SWE | AU | UK | CH | NL | RU | JP | HU | NO |
|---|---|---|---|---|---|---|---|---|---|---|---|---|---|---|---|---|---|
| 1 | USA | 196 | 1 | 0.89 | **0.91** | 0.75 | -0.01 | 0.79 | **0.94** | **0.96** | **0.98** | **0.82** | **0.95** | **0.84** | **-0.32** | **0.93** | **0.99** |
| 2 | France | 174 | | 1 | 0.57 | **0.83** | 0.21 | **0.98** | **0.88** | **0.90** | **0.95** | **0.87** | **0.86** | **0.94** | **-0.96** | **0.84** | **0.90** |
| 3 | Germany | 132 | | | 1 | **0.82** | 0.11 | 0.61 | **0.92** | **0.94** | **0.97** | **0.84** | **0.85** | 0.53 | **-0.48** | **0.96** | **0.80** |
| 4 | Spain | 116 | | | | 1 | 0.67 | 0.22 | 0.64 | 0.66 | 0.78 | 0.54 | **0.83** | 0.57 | **-0.88** | -0.13 | **0.62** |
| 5 | Italy | 116 | | | | | 1 | 0.53 | 0.20 | 0.17 | 0.01 | 0.28 | 0.19 | 0.94 | -0.30 | -0.61 | -0.19 |
| 6 | Finland | 136 | | | | | | 1 | **0.84** | **0.82** | 0.75 | **0.80** | 0.66 | **0.84** | **-0.95** | 0.66 | 0.94 |
| 7 | Sweden | 124 | | | | | | | 1 | **0.99** | **0.97** | **0.95** | **0.87** | **0.84** | **-0.86** | **0.93** | **0.99** |
| 8 | Australia | 195 | | | | | | | | 1 | **0.98** | **0.91** | **0.85** | **0.83** | -0.32 | **0.87** | **0.99** |
| 9 | UK | 195 | | | | | | | | | 1 | **0.86** | **0.92** | **0.83** | -0.27 | **0.90** | **0.97** |
| 10 | CH | 176 | | | | | | | | | | 1 | 0.74 | **0.83** | -0.53 | 0.76 | **0.90** |
| 11 | NL | 112 | | | | | | | | | | | 1 | 0.64 | **-0.84** | **0.91** | **0.82** |
| 12 | Russia | 21 | | | | | | | | | | | | 1 | **-0.92** | 0.61 | **0.86** |
| 13 | Japan | 46 | | | | | | | | | | | | | 1 | 0.17 | -0.52 |
| 14 | Hungary | 67 | | | | | | | | | | | | | | 1 | **0.85** |
| 15 | Norway | 184 | | | | | | | | | | | | | | | 1 |

*Table 6: Correlation coefficients of House Price Indices between countries for 10 countries. Values larger than 0.80 are highlighted in red and italics. UK, CH and NL stand for United Kingdom, Switzerland and Netherlands, respectively. Four datasets (Italy, Finland, Japan, and Spain) correlate less with other price series. UK, Sweden, Australia, and the Netherlands show the strongest correlation with prices in the USA. Russia and Japan only publish yearly data for their Gross Domestic Product statistics. In the rest of the article, those two countries are excluded from the pool of countries modelled.*

## 3.3. Modelling house prices in 15 countries

As many of the price time series are well correlated, and buyers/sellers are not in contact together and likely do not contribute to others local economies, one variable should contribute to all markets to sustain similar price increases. One of the candidates is, of course, LIBOR, but as it has been suggested that LIBOR was manipulated and it was indeed discontinued from Q4 2016 in the FRED database, I have chosen to use the US Treasury rate 10y, which matches both maturity of the investment hedge of banks when pricing the transaction. For the sake of clarity, and for the rest of the article, I shall only display model results for four countries that already experienced crises (USA, UK, Switzerland, and France), and which, in my view, represent the widest variety of contractual setups ( Table 7). All figures from other countries are available in supplementary materials. I have completed a series of models, which are listed in Table 1 and in Table 15. The figures showing the results for the other countries are available in the appendix.

|  | Amortization | Fixed rates? | Source |
|---|---|---|---|
| USA | Yes, continuous | Possible (ARM, 10%) | Link |
| France | Yes, continuous | Yes, variable 30% | Link |
| UK | Yes | Possible, 26% | Link |
| Switzerland | At least 35% after 10-15 yrs, otherwise discontinuous. | Yes, LIBOR for 10-15%, ARM 5% | Link, Link2 |

*Table 7: Broad characteristics of the market. ARMS: Adjustable Rate Mortgage.*

### 3.3.1. A model with three parameters (3-param; CPI, GDP and TR)

I completed the first series of models based on consumer price index (CPI), Gross Domestic Product (GDP), and treasury rates (10yr) data only (

Table 1). The advantage of this model series is that data is freely available in many countries and some of those data go up until 1970s. One disadvantage is that the dataset is smaller, which may impact the quality of the results. The modeling strategy is the same as all the other models (previous and future) presented in this study. I have tested two algorithms in parallel: kNN and tree-bagging. Those approaches were chosen as they are popular with many, and they were able to produce regression. The tree-bagging technique was chosen over others because it is less prone to overfitting than kNN, and the models produced are thought to be easier to explain than those built with other approaches. kNN is well known for performing with small datasets and is easy to implement. Both approach solutions are shown with distinctive colors and curves showing residual values in each figure. When uncertainty; I conclude that a model is not sufficiently robust because it is of the same magnitude as the difference between price series of various origins or between prices of different assets (houses vs flats). In total, five countries' models show an RMS > 15%: (USA, Sweden, Australia, Netherland and Norway; Table 8). Two observations can be noted from this set of third models. First, the fit with the data is good as the long-term features of the price time series are recovered overall. Second, from 2015, the models do not explain the price increase for most countries. I note that France and UK models performs slightly better than USA and Switzerland where loans are not always amortized. The amortization allows damping of price changes following sudden change of fiscal and economical policies. Two explanations may come from the economic perspective: negative interest rates and (mostly for Europe) non-conventional support program of the European Central Bank (e.g., ECB PSPP[7] and ECB PEPP[8]).

---

[7] "PSPP" stands for Public sector purchase program". As specified on the ECB website, the securities included in the PSPP include nominal and inflation-linked central government bonds and bonds issued by recognised agencies, regional and local governments, international organisations and multilateral development banks located in the euro area.

[8] PEPP stands for Pandemic Emergency Purchase Programme. Description of the PEPP program can be found here.

I know those two effects may be interlinked, so I modeled those effects separately. For this model, like for some of the following, I have also computed models after differentiating HPI over 12 quarters (Table 1).

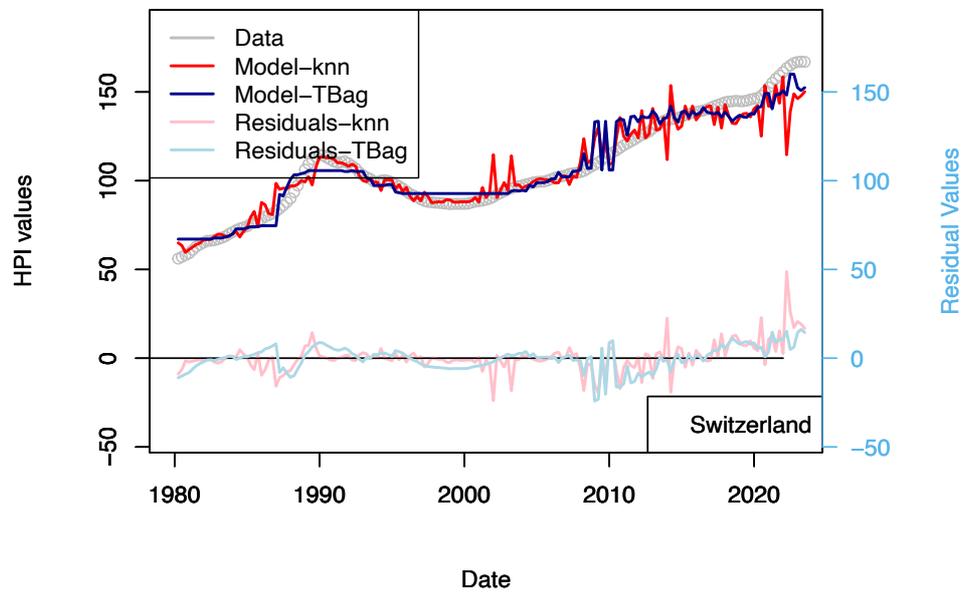

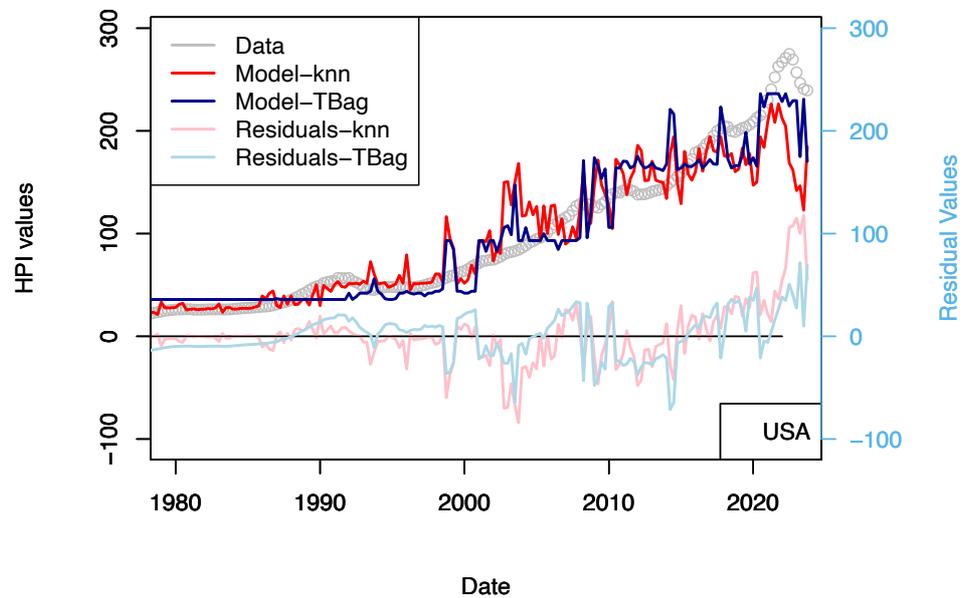

*Figure 4: "3-param" models. For the sake of presentation, only four countries in the 15 time-series modelled are shown in this figure (see the following figure for other results). As explained in the text most models of this series do not predict an increase of price from 2015 nor during the period following the COVID period.*

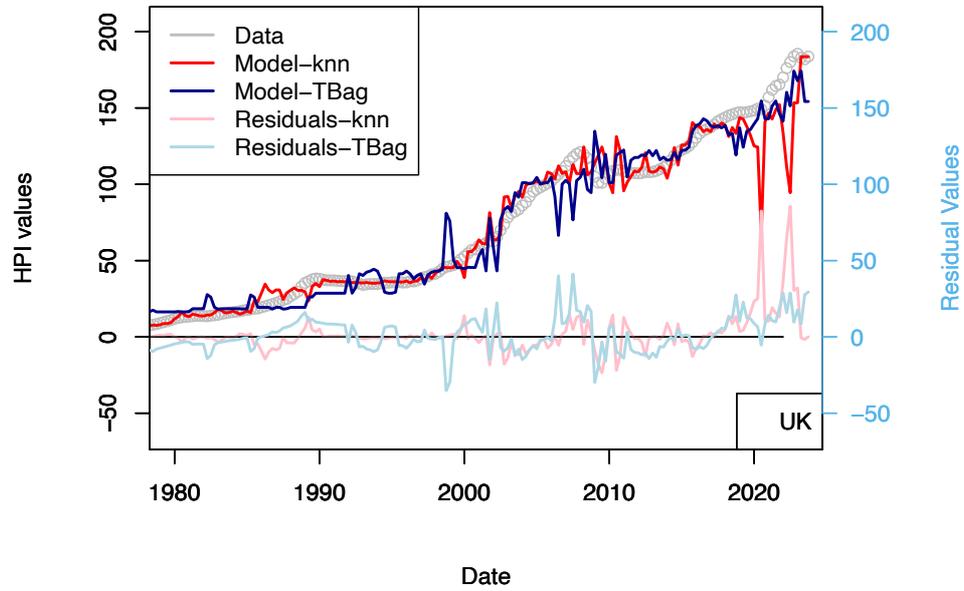

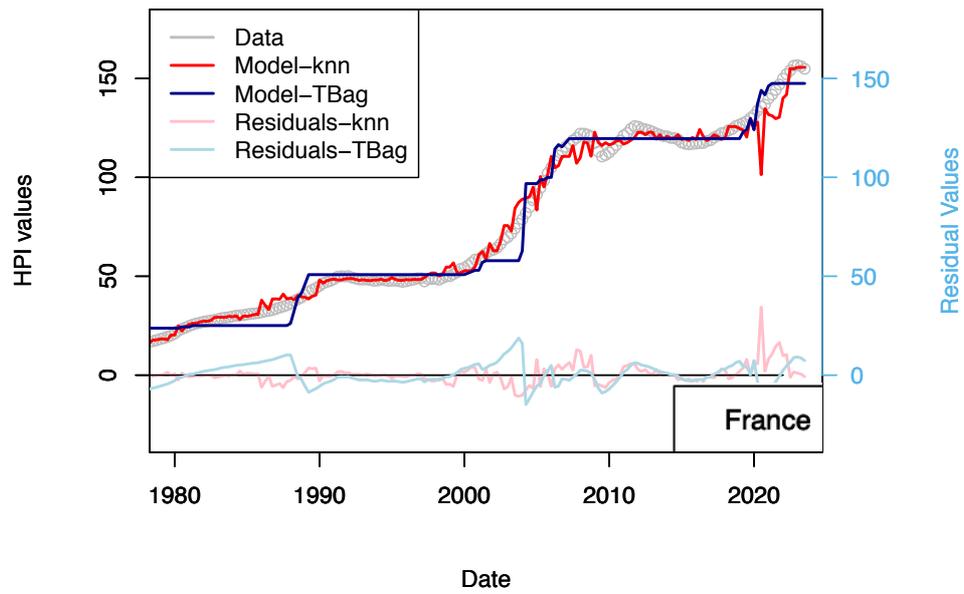

*Figure 4 (continued): "3-param" models. For presentation, only four countries of the 15 time series modeled are shown in this figure (see supplementary materials for other results).*

| Countries | M_COR | S_COR | M_RMS | S_RMS | SD | M_MAE | M_MAPE | M_CHIp | M_CHIs |
|---|---|---|---|---|---|---|---|---|---|
| USA | 0.949 | 0.002 | 22.405 | 0.383 | 31.982 | 17.575 | 0.228 | 0.261 | 0.000 |
| France | 0.959 | 0.001 | 11.637 | 0.167 | 0.295 | 7.946 | 0.102 | 0.183 | 0.007 |
| Germany | 0.924 | 0.004 | 11.087 | 0.333 | 3.415 | 6.945 | 0.055 | 0.316 | 0.000 |
| Spain | 0.882 | 0.005 | 11.404 | 0.231 | 2.683 | 7.658 | 0.098 | 0.173 | 0.067 |
| Italy | 0.865 | 0.004 | 9.298 | 0.126 | 2.554 | 6.349 | 0.070 | 0.189 | 0.034 |
| Finland | 0.956 | 0.001 | 10.143 | 0.121 | 2.572 | 7.552 | 0.095 | 0.217 | 0.004 |
| Sweden | 0.956 | 0.003 | 19.687 | 0.661 | 20.861 | 15.148 | 0.125 | 0.166 | 0.066 |
| Australia | 0.956 | 0.002 | 21.044 | 0.433 | 14.522 | 15.866 | 0.247 | 0.215 | 0.000 |
| UK | 0.976 | 0.001 | 11.617 | 0.313 | 12.483 | 8.950 | 0.193 | 0.181 | 0.004 |
| Switzerland | 0.950 | 0.002 | 8.638 | 0.160 | 6.624 | 7.068 | 0.066 | 0.203 | 0.002 |
| Netherland | 0.865 | 0.011 | 16.400 | 0.514 | 17.635 | 12.710 | 0.123 | 0.160 | 0.131 |
| Norway | 0.965 | 0.001 | 18.465 | 0.339 | NA | 14.244 | 0.178 | 0.149 | 0.037 |

*Table 8: Statistics of the data fit performance of the 3-parameters models (tree-bag strategy). HPI values are in nominal form. COR stands for Correlation, RMS for Root mean squared, MAE for mean absolute error , MAPE for mean absolute percentage error. CHIp and CHIs are the parameter and statistic of the Chi2.*

### 3.3.2. Models with three parameters (CPI, GDP and TR) plus central bank rates: IR and LIR models

Having shown that the prices of real-estate could be successfully interlinked between countries (the best example of such statement being the propagation of the 2008 house price drops from USA to other countries), I have completed two sets of computations. The first set of models comprises those for which US rates applied to all countries, the second set uses the local national central bank rates. A second set of models (Australia, Switzerland, and European countries) use local central bank rates and not the USA ones. HPI values are kept in their nominal form.

Overall, the inclusion of additional data was expected to have a positive impact, and this is well observed here (Table 9). However, the addition of the central bank rates data has different impact. For France, the impact is minimal. This could be explained by the fact that most of the mortgages loans in France are locked in (fixed rates) and are amortized. We can assume that the number of variable rate loans is not significant to cause a drop of prices (limited number of potential forced sales). A change in the central bank rates is only impacting new investors (the youngest of the borrowers) and older investors if rates used in their contracts are variable. A change or rate is only expressed through prices when margin calls cannot be met (prices observed are not a direct function of discrete pockets of liquidity stress). The impact on older borrowers is only visible if their income is decreased or they cannot meet margin calls.

In the USA and UK, the improvement is minimal, suggesting either inflation is collinear with central bank rates, or, for the most recent years, arbitrage to enter a real-estate transaction is partially driven by central bank policies. In Switzerland, we can find a slight improvement over the most recent years (> 2015).

Using national central bank rates leads to the same conclusions: the fit to the data is not significantly improved for most recent observations ( Figure 6). Finally, those data were redundant with the treasury rate 10 yr, which calibrates the rates worldwide when financing is mobilized or when securitization of the position is done to unload the lenders' banking books.

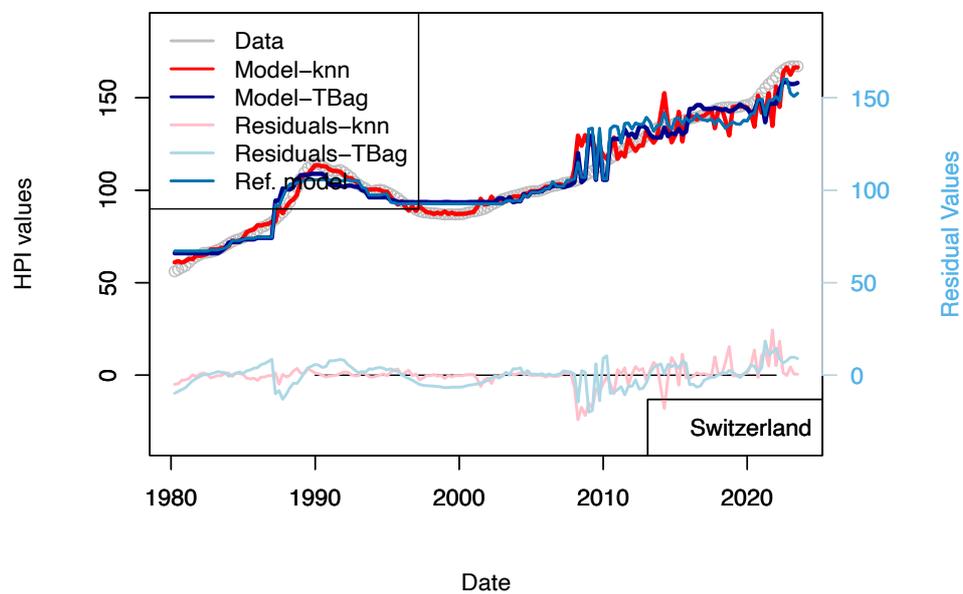

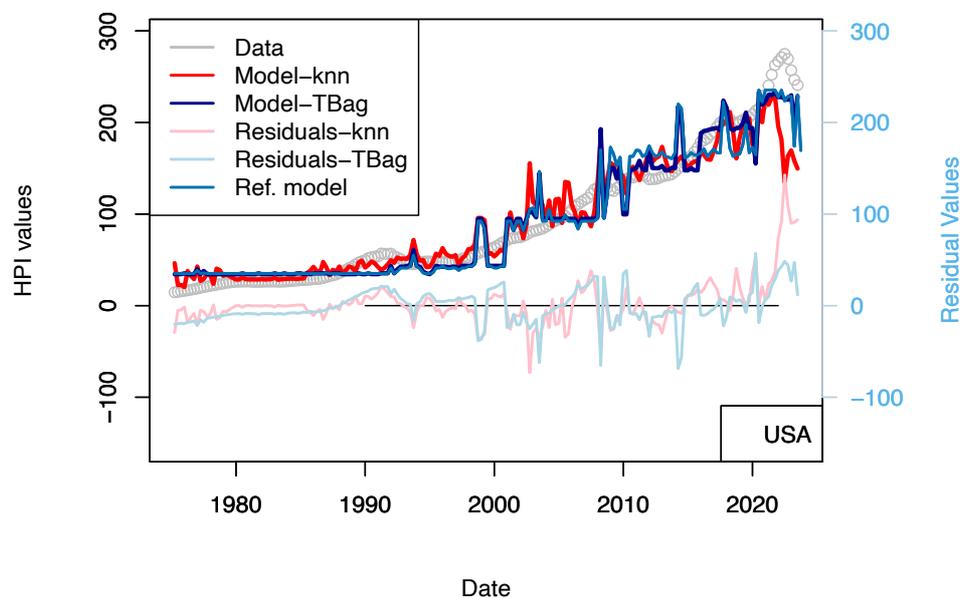

*Figure 5: "Central bank IR" models. Comparison of models including (or not) the US FED interest rates to all countries. Parameters used here are CPI, GDP, and Treasury rates (10 yr), and for one of the two models, the Swiss National Bank (SNB) interest rates. The price history modelled without using the central bank rates is shown using a blue curve. All models benefit from including the interest rates (see all orange curves).*

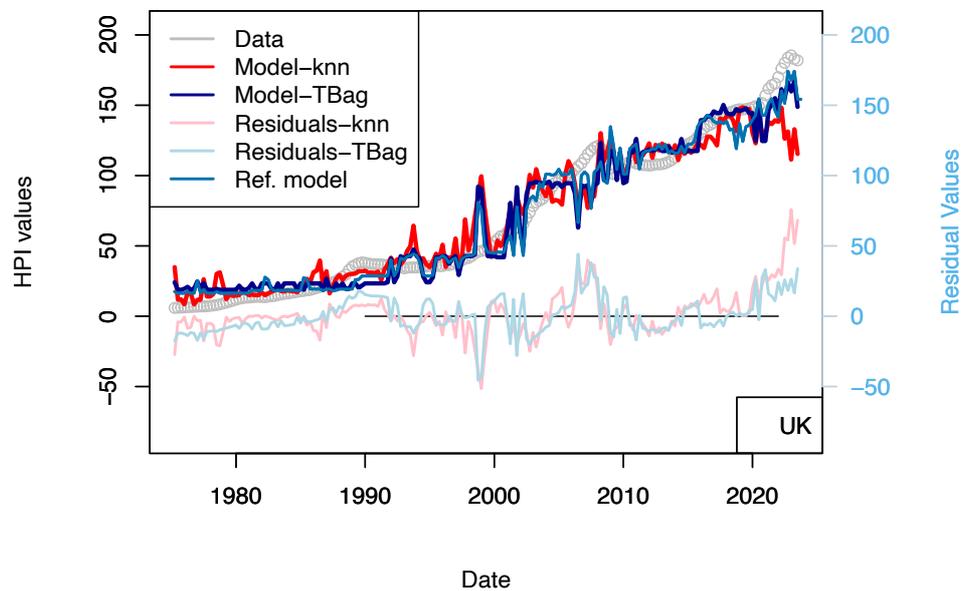

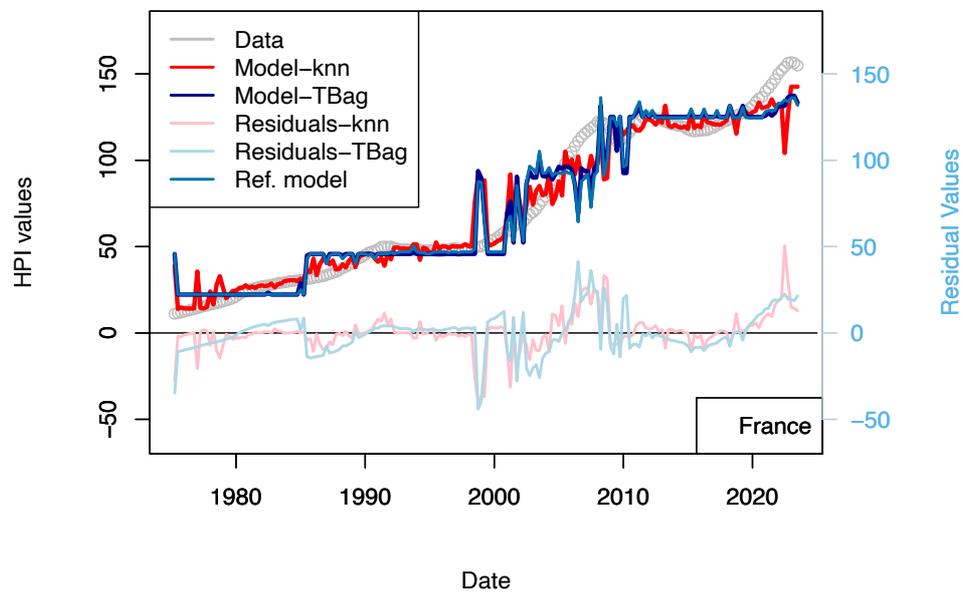

*Figure 5 (continued):* "Central bank IR" models. Comparison of models including (or not) the US FED interest rates to all countries. Parameters used here are CPI, GDP, and Treasury rates (10 yr), and for one of the two models, the Swiss National Bank (SNB) interest rates. The price history modelled without using the central bank rates is shown using a blue curve. All models benefit from including the interest rates (see all orange curves).

| Countries | M_COR | S_COR | M_RMS | S_RMS | SD | M_MAE | M_MAPE | M_CHIp | M_CHIs |
|---|---|---|---|---|---|---|---|---|---|
| USA | 0.963 | 0.002 | 18.871 | 0.443 | 18.999 | 14.092 | 0.204 | 0.262 | 0.000 |
| France | 0.960 | 0.001 | 11.550 | 0.183 | 3.310 | 7.964 | 0.105 | 0.186 | 0.006 |
| Germany | 0.917 | 0.008 | 11.457 | 0.456 | 8.660 | 7.123 | 0.055 | 0.353 | 0.000 |
| Spain | 0.890 | 0.005 | 11.144 | 0.240 | 3.178 | 7.255 | 0.094 | 0.135 | 0.255 |
| Italy | 0.893 | 0.005 | 8.431 | 0.155 | 3.264 | 5.405 | 0.061 | 0.198 | 0.025 |
| Finland | 0.954 | 0.001 | 10.373 | 0.131 | 3.170 | 7.729 | 0.097 | 0.212 | 0.005 |
| Sweden | 0.953 | 0.002 | 20.070 | 0.495 | 9.510 | 15.929 | 0.135 | 0.167 | 0.066 |
| Australia | 0.965 | 0.002 | 18.624 | 0.430 | 13.469 | 13.785 | 0.233 | 0.214 | 0.000 |
| UK | 0.968 | 0.001 | 13.066 | 0.210 | 7.663 | 10.131 | 0.240 | 0.216 | 0.000 |
| Switzerland | 0.961 | 0.001 | 7.472 | 0.126 | 2.100 | 5.972 | 0.057 | 0.182 | 0.007 |
| Netherland | 0.879 | 0.008 | 14.983 | 0.409 | 6.807 | 11.102 | 0.110 | 0.166 | 0.112 |
| Norway | 0.994 | 0.000 | 7.399 | 0.269 | NA | 5.752 | 0.084 | 0.134 | 0.081 |

*Table 9: Statistics of the data fit performance of the IR models (tree-bagging strategy). COR stands for Correlation, RMS for Root mean squared, MAE for mean absolute error, MAPE for mean absolute percentage error. CHIp and CHIs are the parameter and statistic of the Chi2.*

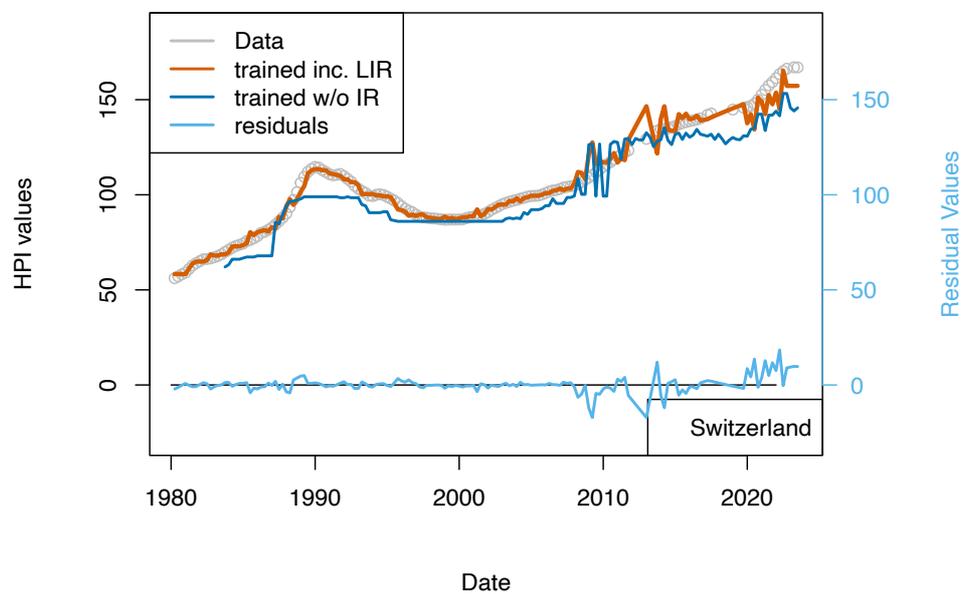

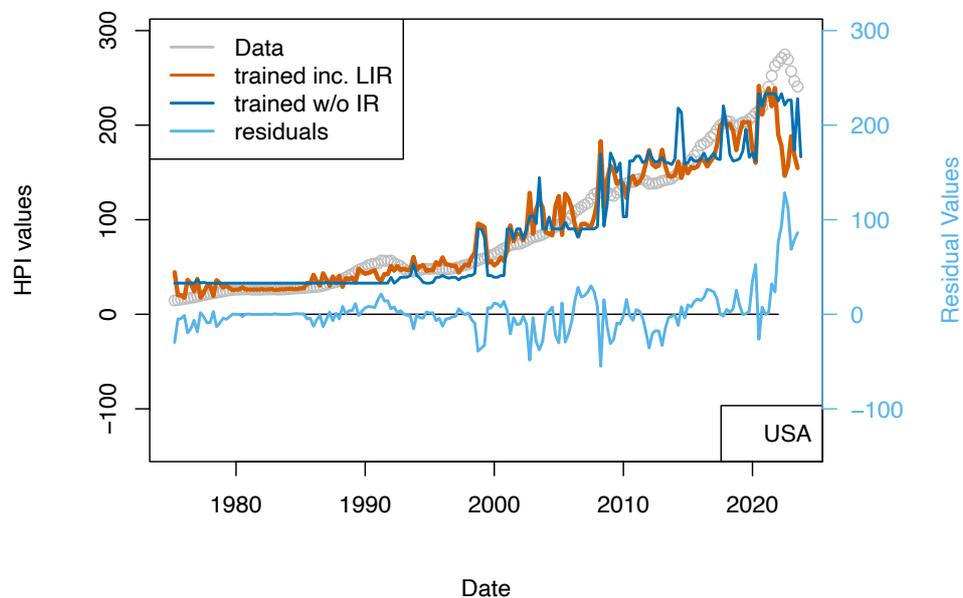

*Figure 6: "LIR" model. Comparison of models including (or not) the US FED interest rates to all countries, unless some local rates are available (CH, EU rates). Parameters used here are CPI, GDP, and Treasury rates (10 yr), and for one of the two models, the Swiss National Bank (SNB) interest rates. The price history modelled without using the central bank rates is shown using a blue curve. All models benefit from including the interest rates (see all orange curves).*

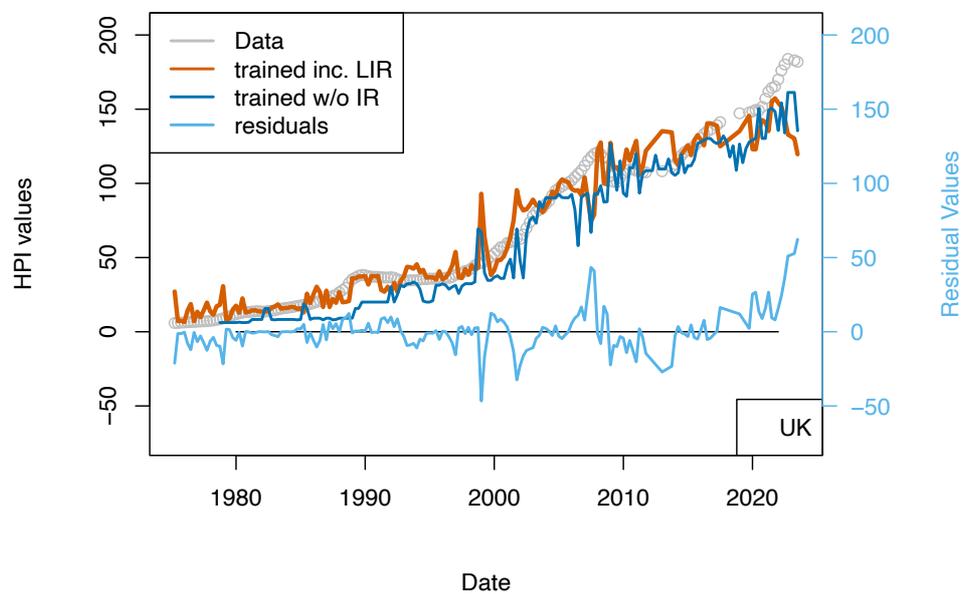

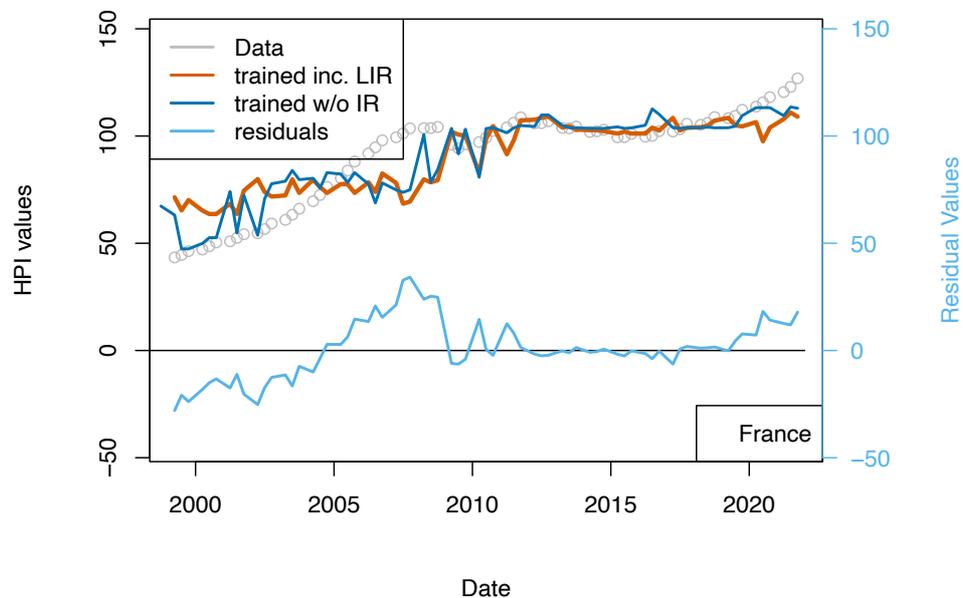

*Figure 6 (continued): "LIR" model. Comparison of models including (or not) the US FED interest rates to all countries unless some local rates are available (CH, EU rates). Parameters used here CPI, GDP, Treasury rates (10yr), and for one of the two models, interest rates of Swiss National Bank (SNB). The price history modelled without using the central bank rates is shown using a blue curve. All models benefit from the inclusion of the interest rates (see all orange curves).*

| Countries | M_COR | S_COR | M_RMS | S_RMS | SD | M_MAE | M_MAPE | M_CHIp | M_CHIs |
|---|---|---|---|---|---|---|---|---|---|
| USA | 0.961 | 0.002 | 19.323 | 0.400 | 18.720 | 14.538 | 0.214 | 0.263 | 0.000 |
| France | 0.897 | 0.004 | 12.552 | 0.191 | 6.631 | 9.746 | 0.098 | 0.193 | 0.061 |
| Germany | 0.864 | 0.011 | 15.170 | 0.558 | 5.149 | 8.765 | 0.065 | 0.261 | 0.003 |
| Spain | 0.850 | 0.010 | 10.154 | 0.234 | 3.019 | 6.801 | 0.083 | 0.163 | 0.161 |
| Italy | 0.857 | 0.008 | 7.797 | 0.150 | 2.474 | 5.210 | 0.058 | 0.229 | 0.017 |
| Finland | 0.923 | 0.003 | 9.365 | 0.169 | 8.135 | 6.541 | 0.067 | 0.167 | 0.139 |
| Sweden | 0.900 | 0.006 | 24.037 | 0.733 | 53.721 | 16.982 | 0.125 | 0.165 | 0.254 |
| Australia | 0.965 | 0.001 | 17.924 | 0.367 | 32.903 | 12.277 | 0.212 | 0.214 | 0.001 |
| UK | 0.963 | 0.002 | 13.596 | 0.260 | 3.191 | 10.558 | 0.253 | 0.226 | 0.001 |
| Switzerland | 0.975 | 0.003 | 5.885 | 0.304 | 16.130 | 4.332 | 0.041 | 0.163 | 0.031 |
| Netherland | 0.873 | 0.014 | 13.497 | 0.464 | 6.959 | 9.563 | 0.086 | 0.212 | 0.026 |
| Norway | 0.964 | 0.002 | 16.851 | 0.443 | NA | 12.070 | 0.123 | 0.264 | 0.001 |

*Table 10: Statistics of the data fit performance of the LIR (local interest rates) models (tree-bagging strategy). COR stands for Correlation, RMS for Root mean squared, MAE for mean absolute error, MAPE for mean absolute percentage error. CHIp and CHIs are the parameter and statistic of the Chi2. Models with RMS<20 and MAPE less than 20% (bold) are considered as performing.*

### 3.3.3. Models including ECB/FED asset sizes

The previous models including central bank interest rates are not sufficient for explaining the recent rise in prices. However, Quantitative Easing (QE) program of the FED/ECB but also by the PEPP of ECB to support the economy may have provided arbitrage to some investors. Those programs started in 2015 and continued until the end of the COVID-19 pandemic. Those programs are nowadays closed by decreasing the book sizes. ECB and FED support programs aim at providing liquidity to the financial sector by relieving them from low long-term assets and by lowering the coupons of corporate debt securities (by increasing the demand on those assets). Because of the nominal involved, such purchase programs were suspected to have an impact on other markets. However, it has been shown that those may have impacted the equity market variously through time and differently for each crisis [53, 54] likely because of a mismatch of time to maturity. For real-estate however, maturities (2, 5 and up to 30) and return to investments (3-5%) are similar, and some investors may then be provided an investment arbitrage. In other words, some investors may be interested in investing in real-estate securities than in corporate or sovereign debt.

Regarding the impact of PSPP/PEPP, it is interesting to note that announcement of those programs stabilizes markets before transactions are executed and cost realized [55]. For those programs, amounts are announced one or two semesters beforehand. Impact is therefore almost immediate. For this reason I chose to use the nominal of ECB asset and not its monthly purchase. I first test the effect of the ECB program (PSPP, PEPP) and then those combined from FED programs (QE, QTs). In any case the collinearity with the price level is low as 1) these programs do not cover the complete period of observation (zeros values are available for when asset size was zero or before ECB existed), and 2) those are being decommissioned.

| Countries | M_COR | S_COR | M_RMS | S_RMS | SD | M_MAE | M_MAPE | M_CHIp | M_CHIs |
|---|---|---|---|---|---|---|---|---|---|
| USA | 0.983 | 0.001 | 12.937 | 0.232 | 3.121 | 10.908 | 0.202 | 0.265 | 0.001 |
| France | 0.993 | 0.001 | 5.019 | 0.176 | 0.829 | 3.787 | 0.059 | 0.159 | 0.030 |
| Germany | 0.985 | 0.001 | 4.982 | 0.201 | 3.129 | 3.739 | 0.031 | 0.400 | 0.001 |
| Spain | 0.980 | 0.002 | 4.806 | 0.194 | 1.286 | 3.582 | 0.047 | 0.181 | 0.045 |
| Italy | 0.978 | 0.002 | 3.905 | 0.136 | 0.198 | 3.151 | 0.036 | 0.243 | 0.002 |
| Finland | 0.985 | 0.001 | 5.923 | 0.136 | 2.450 | 4.831 | 0.064 | 0.204 | 0.008 |
| Sweden | 0.992 | 0.000 | 8.463 | 0.211 | 3.544 | 6.431 | 0.055 | 0.151 | 0.122 |
| Australia | 0.988 | 0.001 | 11.212 | 0.390 | 7.948 | 9.038 | 0.199 | 0.212 | 0.000 |
| UK | 0.984 | 0.001 | 9.399 | 0.217 | 1.096 | 7.665 | 0.204 | 0.191 | 0.002 |
| Switzerland | 0.976 | 0.001 | 5.995 | 0.100 | 0.147 | 4.861 | 0.049 | 0.193 | 0.003 |
| Netherland | 0.981 | 0.002 | 6.220 | 0.348 | 4.405 | 5.218 | 0.054 | 0.207 | 0.018 |
| Norway | 0.990 | 0.001 | 9.631 | 0.310 | NA | 7.755 | 0.133 | 0.200 | 0.003 |

*Table 11: Statistics of the data fit performance of the ECB models (tree-bagging strategy). COR stands for Correlation, RMS for Root mean squared, MAE for mean absolute error , MAPE for mean absolute percentage error. CHIp and CHIs are the parameter and statistic of the Chi2.*

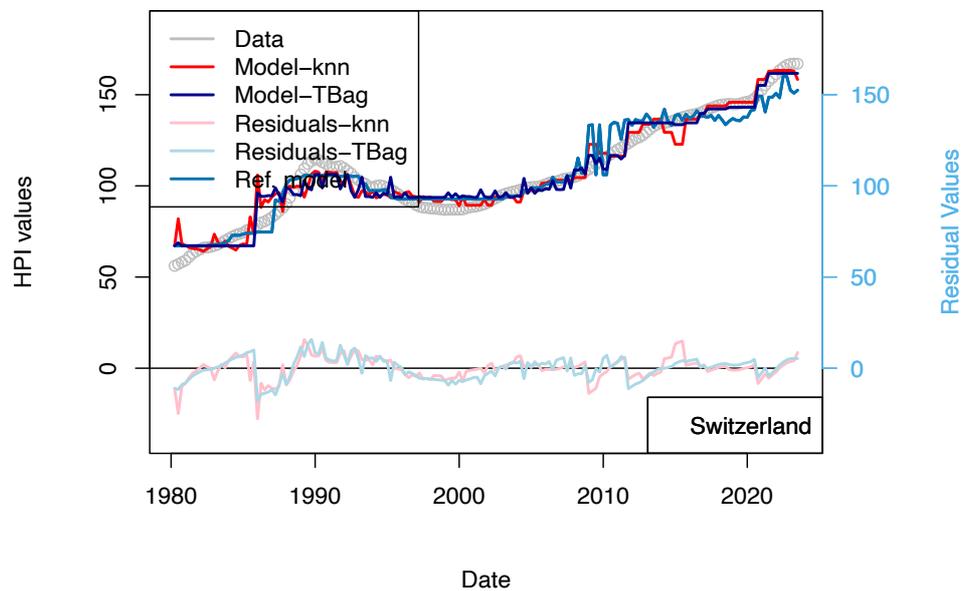

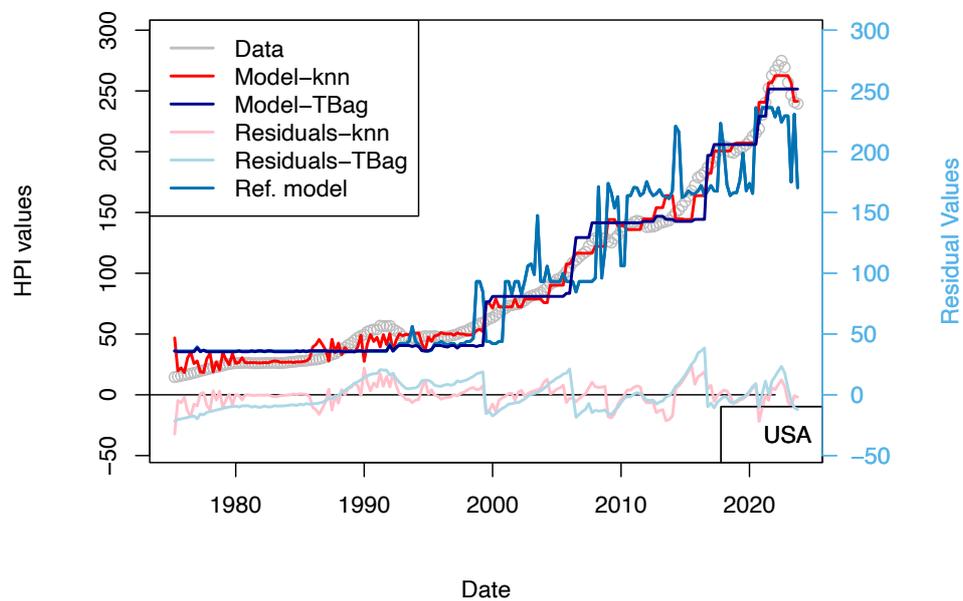

*Figure 7: "ECB" model. With the exception of France all the four time-series fits benefit from the addition of the ECB purchase data. One explanation of this difference could be that in France, the amortization of the asset is complete over the period of the loan. As in combination most loans are on fixed rates, shocks on interest rates are therefore less pronounced impact on the market.*

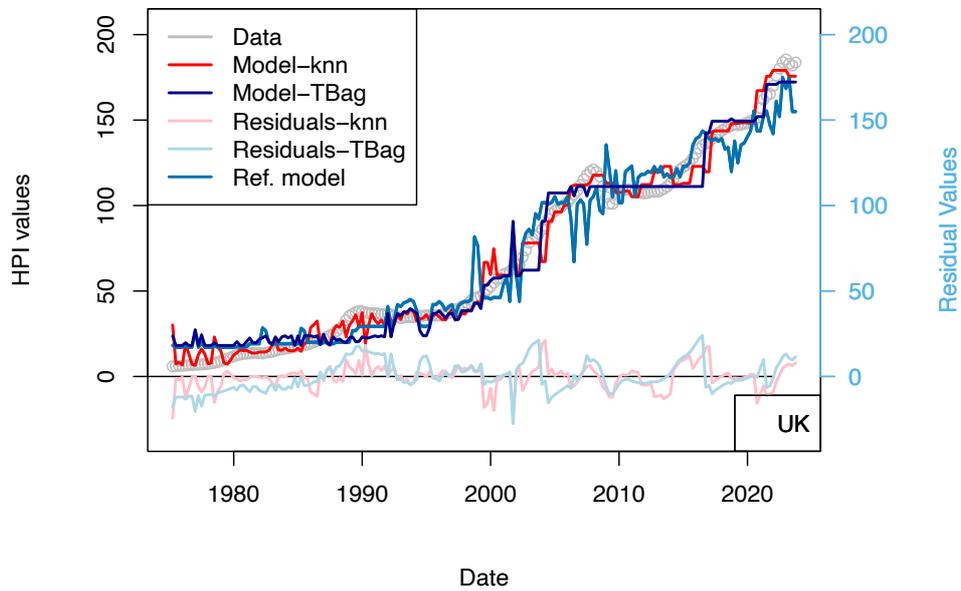

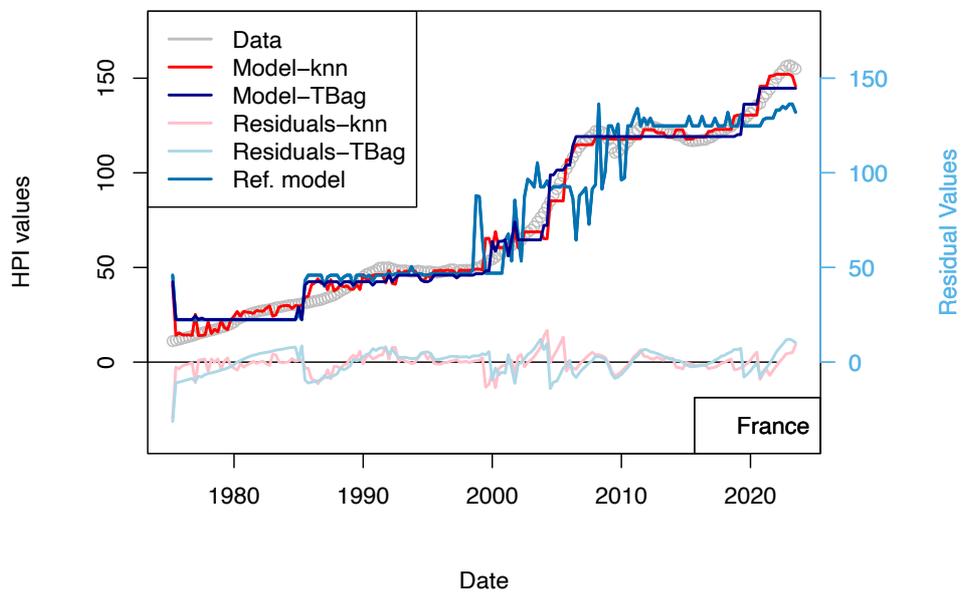

*Figure 7 (continued): "ECB" model. With the exception of France all the four time-series fits benefit from the addition of the ECB purchase data. One explanation of this difference could be that in France, the amortization of the asset is complete over the period of the loan. As in combination most loans are on fixed rates, shocks on interest rates are therefore less pronounced impact on the market.*

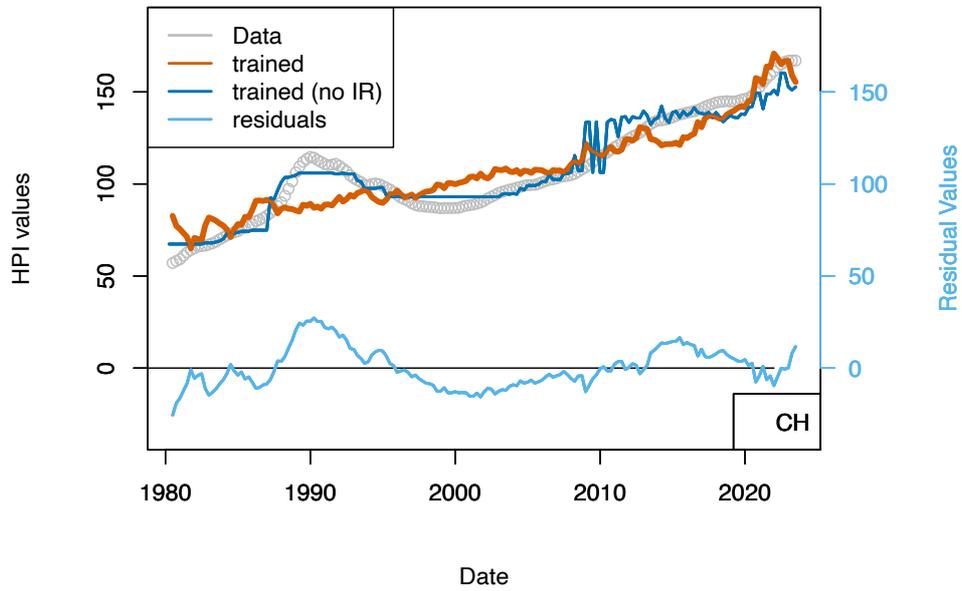

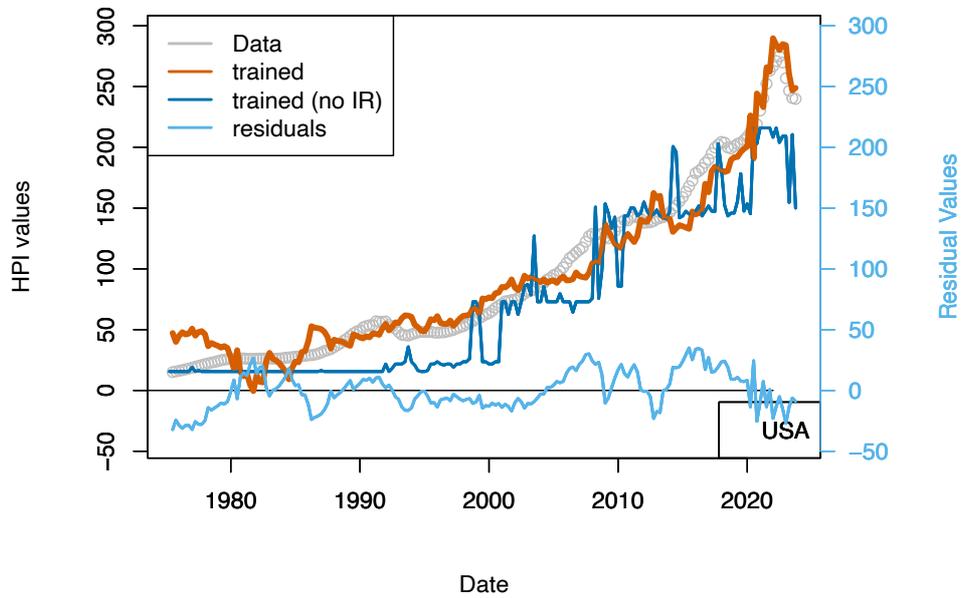

*Figure 8: "ECB/FED" models: Effect of the inclusion of the data of ECB PSPP/PEPP programs and of FED Quantitative Easing (QE) programs (FED total asset book size). The data (grey) and the ECB+FED model (orange) is compared with the 3-param model.*

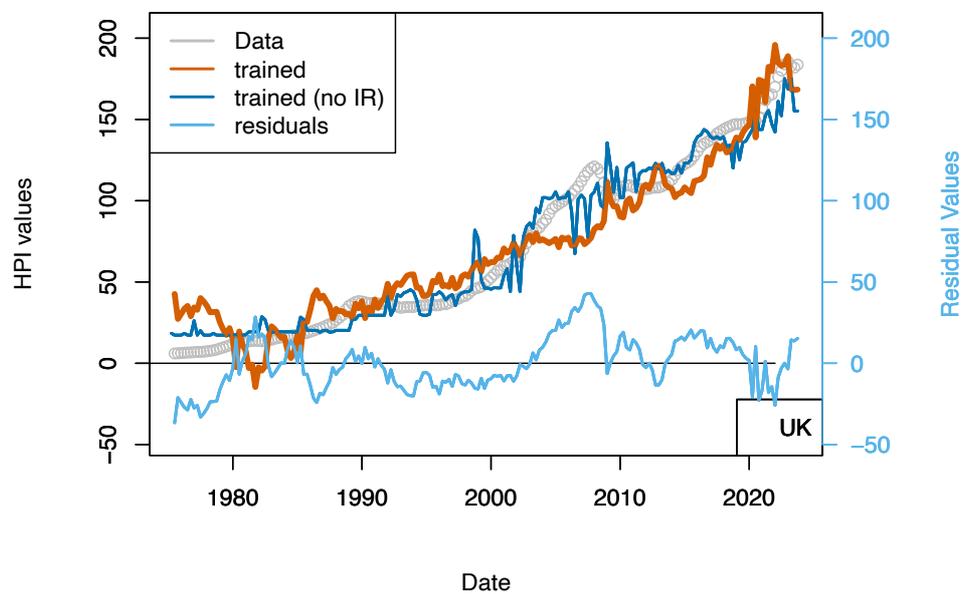

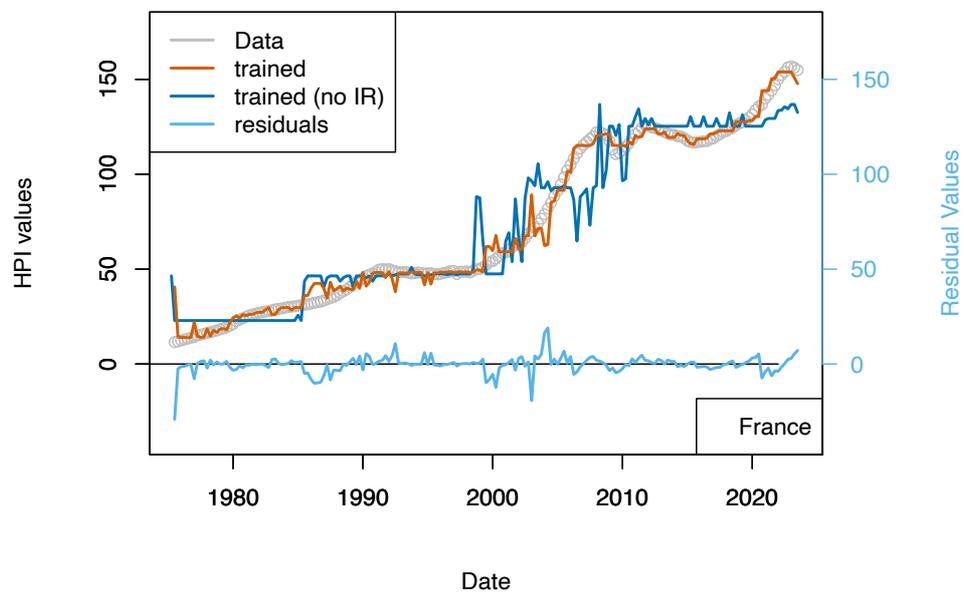

*Figure 8 (continued): "ECB/FED" models: Effect of the inclusion of the data of ECB PSPP/PEPP programs and of FED Quantitative Easing (QE) programs (FED total asset book size). The data (grey) and the ECB+FED model (orange) is compared with the 3-param model.*

For most of the countries, one can see the fit quality is improved for years > 2020 (Table 10). The effect is displayed for both kNN and tree-bag in Figure 7. This indicates that at least the ECB asset book data are not conflicting with the prices of real-estate suggesting a porosity ofsome markets. For the sake of completeness, I have created another model which includes the data of the FED assets book. The fit for US is even further improved, although it is difficult to draw definite concluions from the addition of this dataset.

| Countries | M_COR | S_COR | M_RMS | S_RMS | SD | M_MAE | M_MAPE | M_CHIp | M_CHIs |
|---|---|---|---|---|---|---|---|---|---|
| USA | 0.985 | 0.001 | 12.428 | 0.220 | 1.929 | 10.599 | 0.196 | 0.262 | 0.001 |
| France | 0.993 | 0.001 | 5.009 | 0.193 | 0.828 | 3.758 | 0.060 | 0.165 | 0.025 |
| Germany | 0.980 | 0.002 | 5.642 | 0.230 | 3.827 | 3.979 | 0.032 | 0.381 | 0.001 |
| Spain | 0.983 | 0.002 | 4.435 | 0.187 | 1.146 | 3.404 | 0.044 | 0.182 | 0.044 |
| Italy | 0.974 | 0.002 | 4.216 | 0.130 | 0.434 | 3.500 | 0.040 | 0.246 | 0.002 |
| Finland | 0.987 | 0.001 | 5.468 | 0.116 | 2.450 | 4.490 | 0.060 | 0.232 | 0.002 |
| Sweden | 0.993 | 0.000 | 7.707 | 0.186 | 3.121 | 5.585 | 0.050 | 0.154 | 0.110 |
| Australia | 0.989 | 0.001 | 10.814 | 0.372 | 8.358 | 8.440 | 0.186 | 0.204 | 0.001 |
| UK | 0.984 | 0.001 | 9.234 | 0.214 | 1.346 | 7.465 | 0.208 | 0.196 | 0.002 |
| Switzerland | 0.975 | 0.001 | 6.095 | 0.099 | 0.301 | 5.011 | 0.050 | 0.198 | 0.002 |
| Netherland | 0.980 | 0.003 | 6.371 | 0.383 | 2.271 | 5.420 | 0.056 | 0.203 | 0.023 |
| Norway | 0.995 | 0.000 | 7.000 | 0.291 | NA | 5.523 | 0.089 | 0.141 | 0.058 |

*Table 12: Statistics of the data fit performance of the ECB/FED models (tree-bagging strategy). COR stands for Correlation, RMS for Root mean squared, MAE for mean absolute error , MAPE for mean absolute percentage error. CHIp and CHIs are the parameter and statistic of the Chi2.*

### 3.3.4. The "rent" model (Switzerland)

As some real-estate objects are included into various types of transactions (a same flat can be rented and owned by a private individual or being part of a buy-to-let transaction), I have completed a model for Switzerland for which the rent levels (source: cc-f-05.06.01.17, Office de la Statistique, Suisse) are included. This rent index starts in 1939, therefore it is not limiting the length of the dataset considered so far.

By doing so I take into account that rent and buy markets are connected at least for the same kind of assets. When housing unit production becomes more abundant, then vacancy may increase and rents decrease. When standard of housing improves (insulation, connection to transports, etc.) the rent levels of existing houses may decrease locally. The addition of the rent level (percentage) for each quarter greatly improved the fit of the data (smoother fit). The correlation between modelled and observed prices is > 0.99. The standard deviation of the residual values is of 3.23 and the RMS is of 3.22. After showing that external influence of IR is limited, at least for Switzerland, this last model shows that the rental market influences the property market to some extents. In Switzerland, the rental market is strongly linked with the development of the economy as companies are looking for skilled workers who tend to reside in city center where housing is available. For the purchase market, the link is more complicated to establish. Indeed, the new-comers are kept out of the market for the time needed to build a sufficient down-payment and to secure an income level which qualifies them to satisfy the affordability criteria. While prices and rents continue to increase, it becomes more profitable to complete a transaction of buy-to-let type leading to a concentration of the real-estate to a limited number of players (because of the affordability criteria). In other words, the buy market only benefit of the additional new workers can afford to exit the rent market.

### 3.3.1. Year to Year models: moving away from HPI nominal values

In all models, the input parameters are presented in the form of rates. In order to address the concerns of the time dependance of the parameter to predict, I have differentiated the HPI values over 12 quarters (3 years). Such a period is representative of the period considered by a buyer before taking the decision of buying an asset (e.g. "is the market going up or down?", "is it more profitable to rent rather than buying?"). In these models, the collinearity of data is reduced to its maximum. New versions of the models 3-param, ECB, Local and "Rents" have

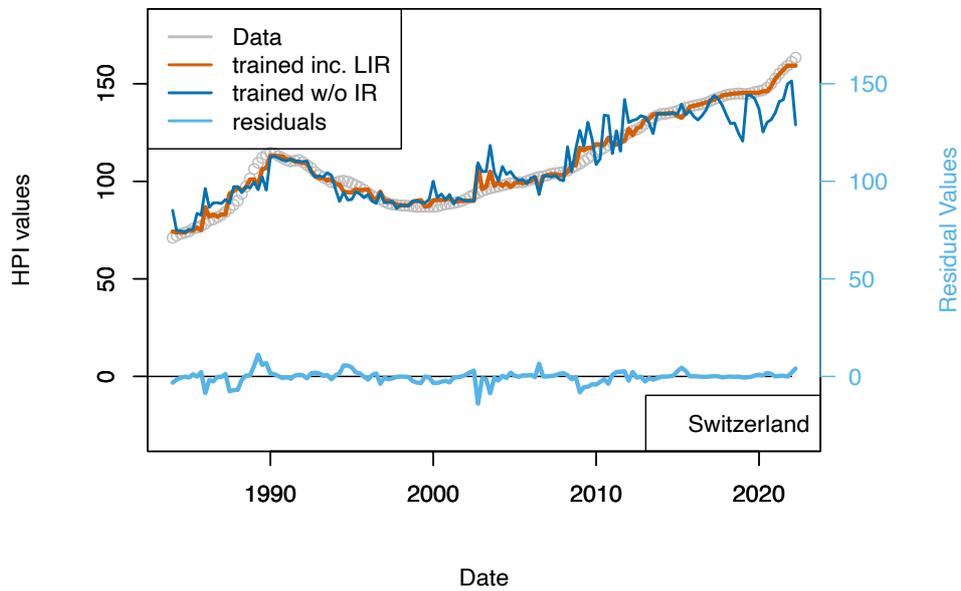

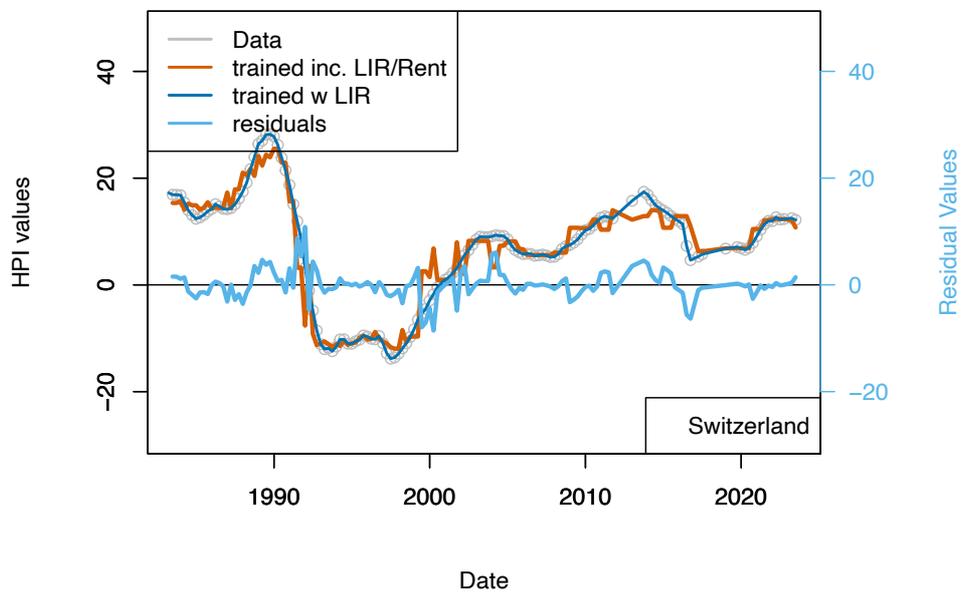

*Figure 9: a) "Rents" model. Results of the dataset including the rent level for Switzerland. Rents levels date back to 1939 (base 100) and have been differentiated. b) "Rent model". Comparison of the model including PSPP, SNB interest rates and rent levels for Switzerland (orange) with the model including only TR, IR, GDP and inflation (dark blue). This figure suggests some level of overfitting of the first versions of the model.*

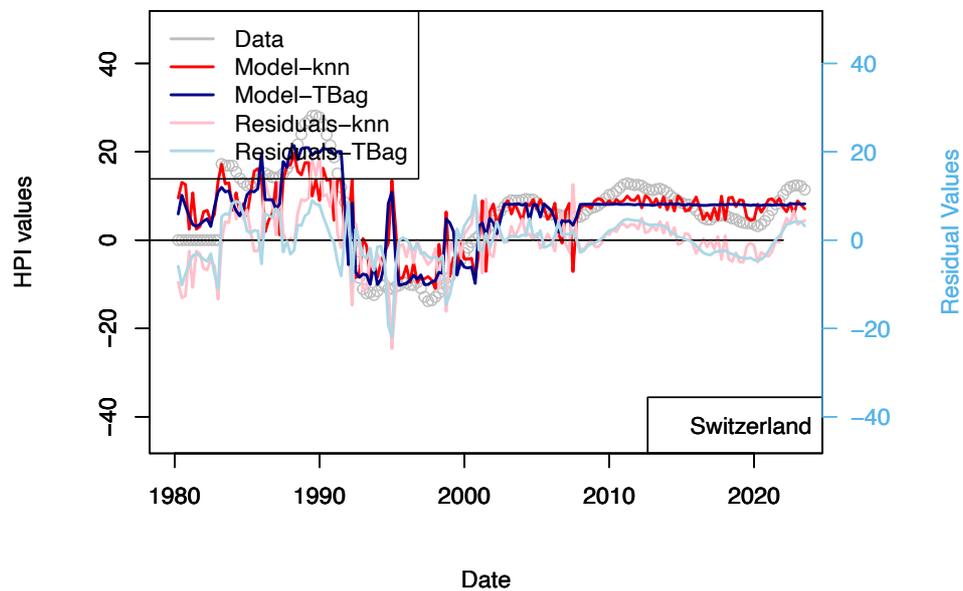

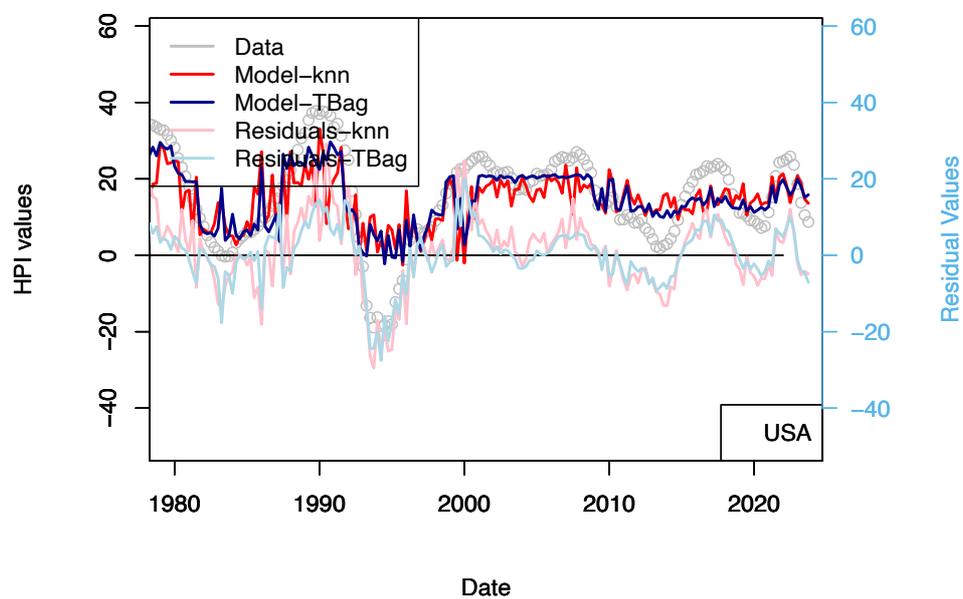

Figure 10: *"3-param 1yr" model. Input data are the same as those used for the model presented in* Figure 4. *HPI are differentiated over 12 quarters*.

been completed. The model performances of these models are displayed in Table 15 and

Table 16. All these models have a good performance with a RMS < 8%. Some of these models do not include ECB data which indicate that the models including ECB asset data can also perform well.

| Countries | M_COR | S_COR | M_RMS | S_RMS | SD | M_MAE | M_MAPE | M_CHIp | M_CHIs |
|---|---|---|---|---|---|---|---|---|---|
| USA | 0.812 | 0.008 | 7.605 | 0.105 | 4.338 | 5.646 | 1.099 | 0.191 | 0.003 |
| France | 0.838 | 0.005 | 6.019 | 0.066 | 2.383 | 4.712 | 0.992 | 0.240 | 0.000 |
| Germany | 0.872 | 0.004 | 3.840 | 0.055 | 4.925 | 2.809 | 2.197 | 0.240 | 0.001 |
| Spain | 0.873 | 0.005 | 8.328 | 0.135 | 2.025 | 6.427 | 1.233 | 0.142 | 0.193 |
| **Italy** | **0.954** | **0.001** | **10.382** | **0.136** | **2.450** | **7.648** | **0.096** | **0.217** | **0.004** |
| **Finland** | **0.848** | **0.006** | **4.522** | **0.081** | **4.528** | **3.717** | **0.360** | **0.204** | **0.012** |
| **Sweden** | **0.808** | **0.006** | **6.471** | **0.071** | **1.446** | **5.200** | **0.346** | **0.212** | **0.000** |
| Australia | 0.891 | 0.004 | 6.736 | 0.110 | 1.070 | 5.204 | 1.515 | 0.147 | 0.042 |
| UK | 0.887 | 0.004 | 4.483 | 0.066 | 1.508 | 3.328 | 0.565 | 0.280 | 0.000 |
| Switzerland | 0.777 | 0.013 | 9.610 | 0.125 | 1.429 | 8.201 | 4.069 | 0.255 | 0.002 |
| Netherland | 0.707 | 0.012 | 23.067 | 0.290 | 3.350 | 16.908 | 0.838 | 0.571 | 0.001 |
| **Norway** | **NA** | **0.000** | **NA** | **0.000** | **NA** | **14.244** | **0.178** | **0.149** | **0.037** |

*Table 13: Statistics of the data fit performance of the 3-parameters-1yr models (tree-bagging strategy). COR stands for Correlation, RMS for Root mean squared, MAE for mean absolute error, MAPE for mean absolute percentage error. CHIp and CHIs are the parameter and statistic of the Chi2.*

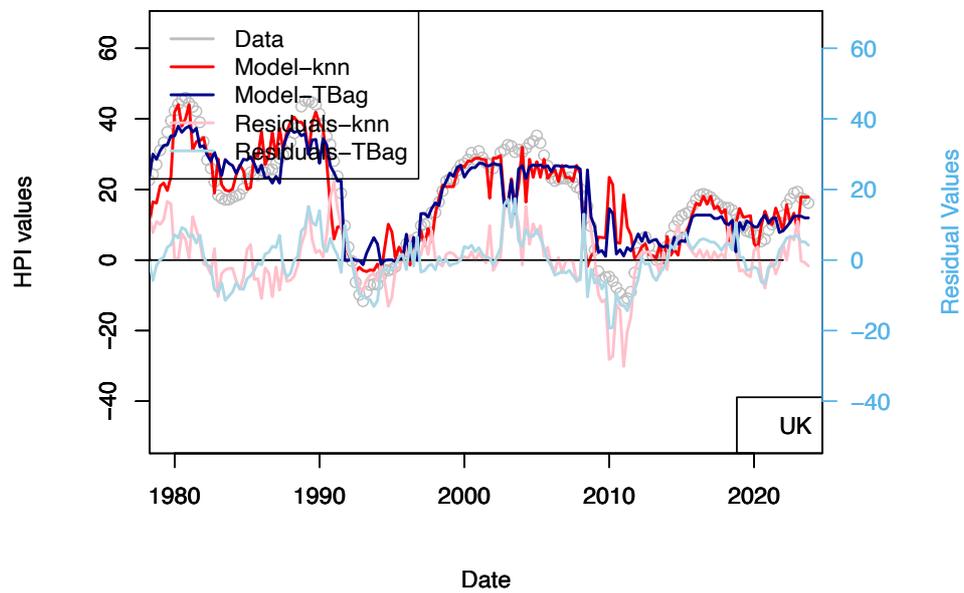

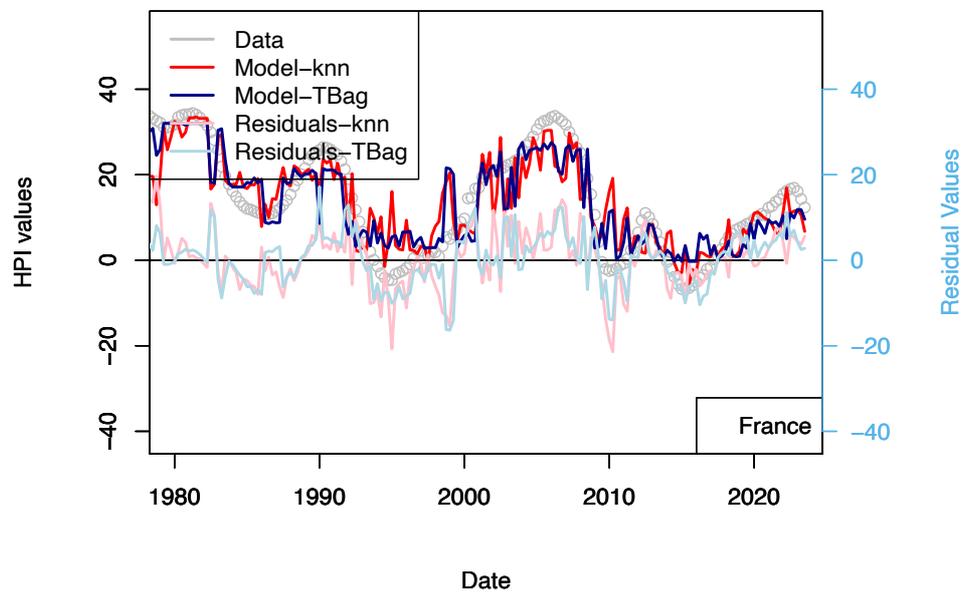

*Figure 10 (continued): "3-param 1yr" model. Input data are the same as those used for the model presented in Figure 5. HPI are differentiated over 12 quarters.*

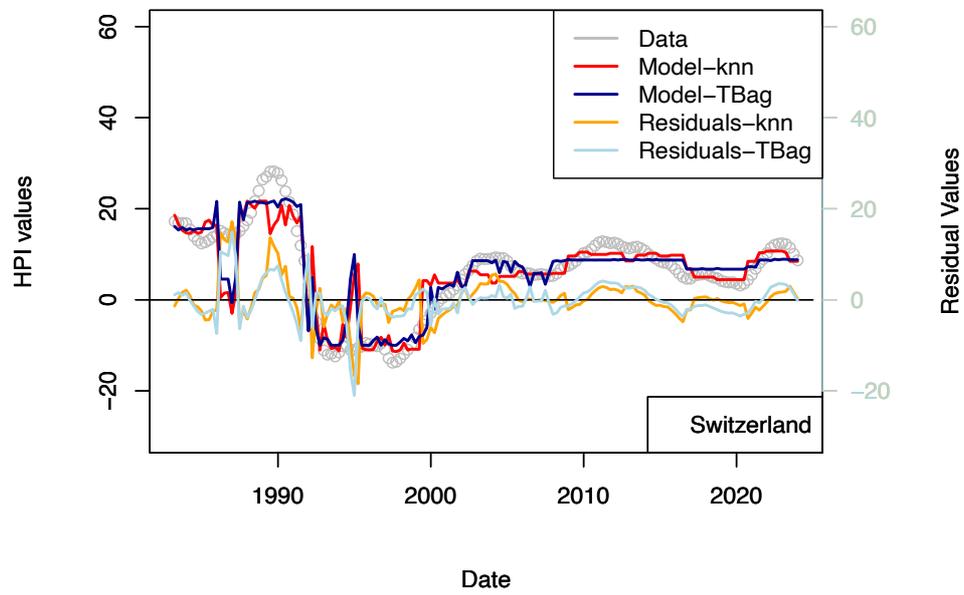

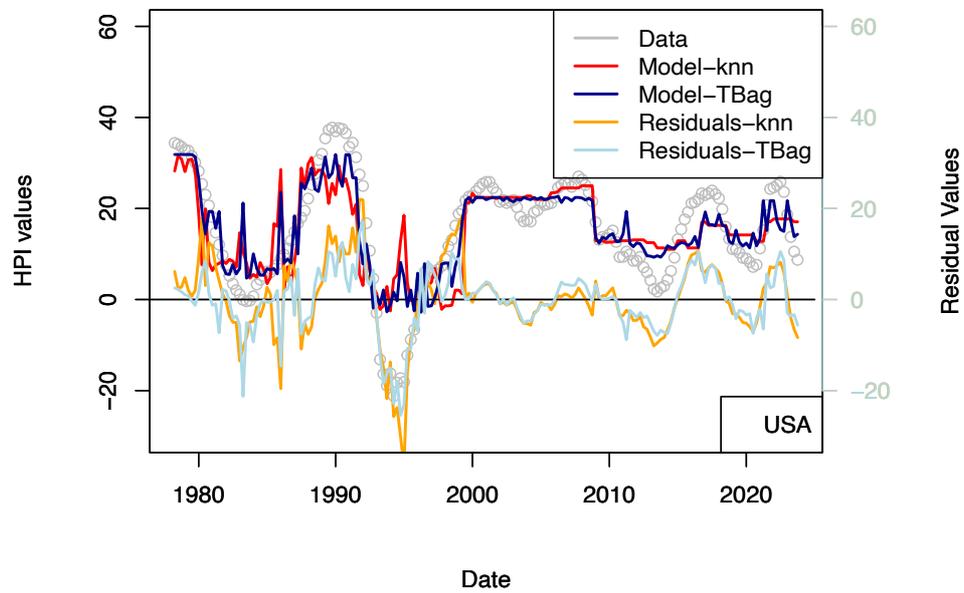

Figure 11: ECB 1yr model. Input data are the same as those used for the model presented in Figure 7. HPI are differentiated over 12 quarters.

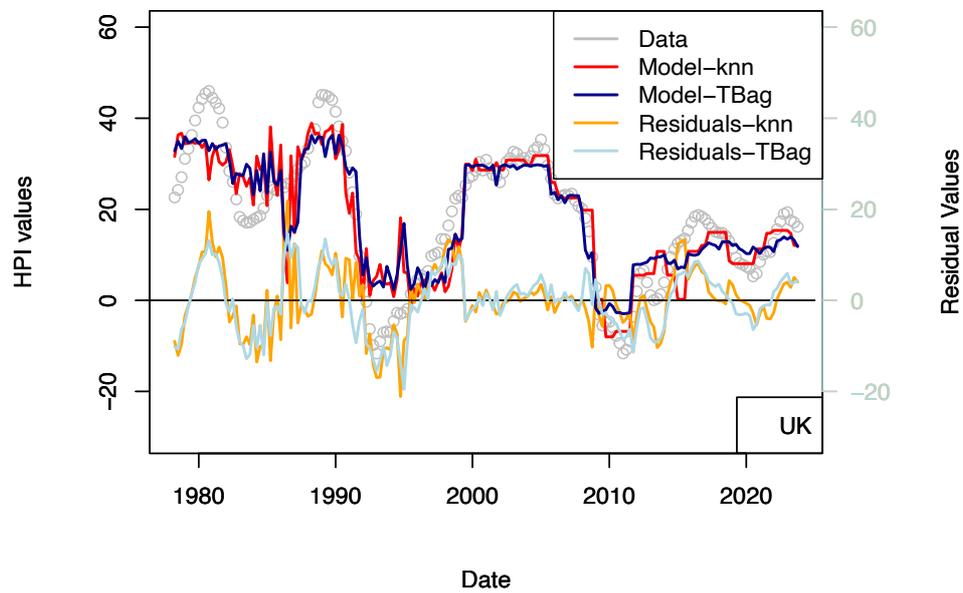

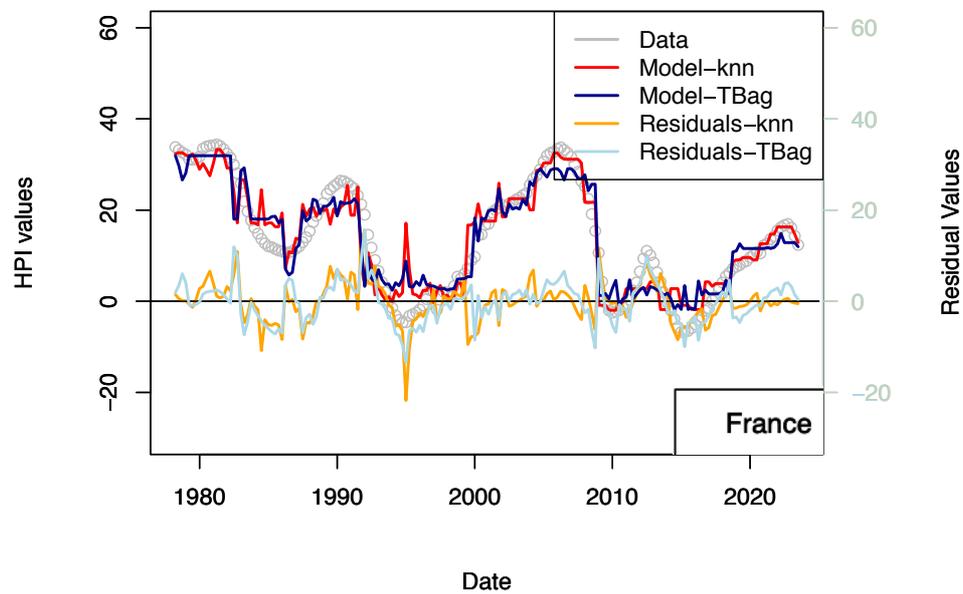

*Figure 11 (continued): ECB 1yr model. Input data are the same as those used for the model presented in Figure 7. HPI are differentiated over 12 quarters.*

| Countries | M_COR | S_COR | M_RMS | S_RMS | SD | M_MAE | M_MAPE | M_CHIp | M_CHIs |
|---|---|---|---|---|---|---|---|---|---|
| USA | 0.850 | 0.007 | 6.680 | 0.121 | 4.901 | 4.828 | 1.230 | 0.159 | 0.023 |
| France | 0.923 | 0.003 | 4.255 | 0.073 | 1.416 | 3.300 | 1.073 | 0.151 | 0.051 |
| Germany | 0.933 | 0.003 | 2.878 | 0.052 | 4.986 | 2.180 | 4.607 | 0.262 | 0.001 |
| Spain | 0.956 | 0.002 | 5.200 | 0.112 | 1.285 | 3.930 | 0.303 | 0.142 | 0.248 |
| Italy | 0.958 | 0.004 | 3.647 | 0.166 | 1.764 | 2.758 | 2.252 | 0.143 | 0.250 |
| Finland | 0.774 | 0.010 | 9.773 | 0.180 | 2.622 | 5.984 | 1.067 | 0.209 | 0.012 |
| Sweden | 0.866 | 0.009 | 3.824 | 0.099 | 3.415 | 3.166 | 0.233 | 0.214 | 0.019 |
| Australia | 0.895 | 0.005 | 4.723 | 0.091 | 1.657 | 3.918 | 0.243 | 0.191 | 0.003 |
| UK | 0.893 | 0.004 | 6.648 | 0.085 | 1.101 | 5.177 | 0.595 | 0.156 | 0.027 |
| Switzerland | 0.914 | 0.003 | 3.935 | 0.059 | 1.401 | 2.742 | 0.417 | 0.212 | 0.002 |
| Netherland | 0.958 | 0.003 | 4.203 | 0.146 | 1.388 | 2.888 | 0.487 | 0.115 | 0.532 |
| Norway | 0.991 | 0.001 | 9.400 | 0.309 | NA | 7.492 | 0.129 | 0.200 | 0.003 |

Table 14: Statistics of the data fit performance of the ECB 1yr models (tree-bagging strategy). COR stands for Correlation, RMS for Root mean squared, MAE for mean absolute error, MAPE for mean absolute percentage error. CHIp and CHIs are the parameter and statistic of the Chi2.

| | Model name | TR | CPI | GDP | Central bank rate | Asset | Additional data, Imp. |
|---|---|---|---|---|---|---|---|
| 1 | 3-param | 100 | >10 | > 10 | - | - | - |
| 2 | 3-param 1yr | 100 | >10 | > 10 | - | - | - |
| 3 | Central bank IR models (US rate) | 100 | >5 | 0 | 75 | - | - |
| 4 | Local Central bank IR models (LIR) | 100 | 93 | 0 | 27.9 | - | - |
| 5 | ECB | 90.9 | 13.4 | 0 | - | 100 | |
| 6 | ECB/FED | 92.9 | >5 | >5 | - | 100 | FED. Rate 24.9 |
| 7 | ECB 1yr | 100 | 5 | 0 | - | 1.7 | - |
| 8 | LOCAL | - | 100 | 22.5 | - | - | Unemployment, 0 |
| 9 | LOCAL 1yr | - | 100 | 43 | | | Joblesss, 0 |
| 10 | Rents | 100 | 0.8 | 0 | - | - | Rent index, 87.0 |
| 11 | Rents 1yr | 78.4 | >1 | 0 | - | - | Rent index, 100 |

Table 15: Variable importance data for the models presented in this paper for Switzerland. The effect of the ECB dataset is visible in the improvement of the model performances.

|   | Model name | Father | TR | CPI | GDP | Central bank rate | Asset | Additional data, Imp. |
|---|---|---|---|---|---|---|---|---|
| 1 | 3-param | - | 100 | 13.4 | 0 | - | - | - |
| 2 | 3-param 1yr | - | 100 | 1.3 | 0 | - | - | - |
| 3 | Central bank IR models (US rate) | 1 | 100 | 12.6 | 0 | 86.2 | - | - |
| 4 | Local Central bank IR models | 1 | 80.8 | 100 | 0 | 56.6 | - | - |
| 5 | ECB | 1 | 90.9 | 13.4 | 0 | - | 100 | |
| 6 | ECB/FED | 1,4 | 76.8 | >5 | 0 | - | 100 | FED. rate >5 |
| 7 | ECB 1yr | 2 | 100 | 5 | 0 | - | 1.7 | - |
| 8 | LOCAL | - | | 100 | 0 | - | - | Unemployment, 22.8 |
| 9 | LOCAL 1yr | | | 100 | 0 | | | 17.6 |
| 10 | Rents | | 77 | >10 | 0 | - | - | Rent index, 100 |
| 11 | Rents 1yr | | 61.5 | >5 | 0 | - | - | Rent index, 100 |

*Table 16: Same as Table 15 for the tree-bag approach.*

## 3.4. Performance tests

Time series used in this study are very smooth. That is, a set of data for a given time is not very different from the previous and the next one. As a consequence, the train/test approach used to estimate the performance of ML models is most of the time successful. A very quick test with introduction of disturbed data (one record multiplied by 5) lead to conclude this approach is not strong enough to determine whether the model is reliable. In the same spirit of what has been done by Dimopoulos and Bakas [56], I have designed more stringent tests and procedures to evaluate the models. In order to get a deeper insight I have completed other tests such as permutation, deletion of data and use of unrelated parameters. First, I completed a permutation exercise which shows that parameters (TR, CPI and GDP) do not have the same influence even when collinearity is not removed. The second test shows that the sole use of local data cannot explain the price increase. At last, I tested the models by quantifying their predictive capabilities and not only from the quality of the data fit.

### 3.4.1. Stability of the residual time-series

In order to confirm a possible time-dependance of the price time-series, I have computed for all models the stability of the residual time-series computed for the tree-bagging model; which I consider as the less stable ones). It is well visible that the models for which HPI has not been differentiated and those not using the ECB are the most variable (models in plain font, p>0.10, Table 17).

| n | Model of residuals series | Statistics | p.value |
|---|---|---|---|
| 1 | Switzerland | -0.70 | 0.97 |
| 2 | **Switzerland 1yr** | **-3.65** | **0.03** |
| 3 | Switzerland IR | -2.13 | 0.52 |
| 4 | Switzerland LIR | -1.45 | 0.81 |
| 5 | **Switzerland ECB** | **-3.64** | **0.03** |
| 6 | **Switzerland ECB_1y** | **-4.25** | **0.01** |
| 7 | France res | -2.26 | 0.47 |
| 8 | **France 1yr** | **-3.58** | **0.04** |
| 9 | France res_IR | -2.13 | 0.52 |
| 10 | France res_LIR | -1.30 | 0.86 |
| 11 | **France ECB** | **-5.73** | **0.01** |
| 12 | **France ECB_1y** | **-3.53** | **0.04** |
| 13 | UK | -3.62 | 0.03 |
| 14 | **UK 1yr** | **-3.45** | **0.05** |
| 15 | **UK IR** | **-3.31** | **0.07** |
| 16 | UK LIR | -1.94 | 0.60 |
| 17 | **UK ECB** | **-5.04** | **0.01** |
| 18 | **UK ECB_1y** | **-4.02** | **0.01** |
| 19 | USA | -2.20 | 0.49 |
| 20 | **USA 1yr** | **-3.94** | **0.01** |
| 21 | USA IR | -3.19 | 0.09 |
| 22 | USA LIR | -1.59 | 0.75 |
| 23 | **USA ECB** | **-4.18** | **0.01** |
| 24 | **USA ECB_1y** | **-4.41** | **0.01** |

*Table 17: Results of the Dickey-Fuller test on the residual series for various models for Switzerland, France, United Kingdom and USA. Models indicated with bold fonts are judged as performant.*

### 3.4.2. Permutations test (model with 3-param).

As the machine-learning algorithms are always assumed to be reliable to model any data using any input parameter, I tried to model the prices using another set of values. I chose to permute original data before modelling prices. In order to make sure the permutation of single parameteris well quantified. I have made three models for which input parameters are permuted independently (

*Figure 12*). As expected, this exercise shows that the input parameter responsible for financing the transaction (TR10yr) is the most important parameter for Switzerland and is the least one. Financing rates impact the pool of people who can afford the transaction. As GDP may the seen as non-impacting on the population buying houses, CPI also plays a role especially to explain short term variations and amplitude changes. As I consider the data at the national level I am not able to detect local effect of GDP like [57] did nor if the number of performing loans are more numerous when economic crisis hits [27]. For the latter effect, it is not clear whether the number of NPL actually has an impact on prices. The conclusion of this test is that the algorithm is not able to model anything using any data and therefore the parameters included in the first models were containing relevant information to price evolution.

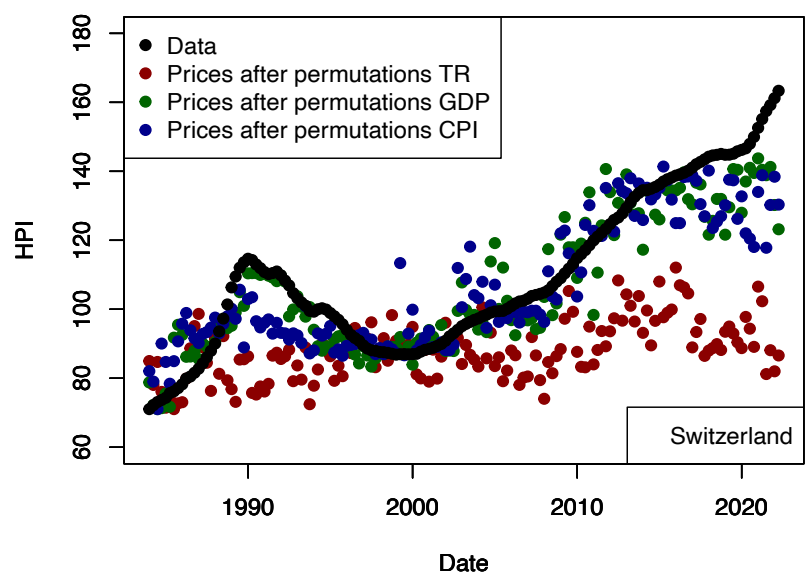

*Figure 12: Comparison of prices for Switzerland (black circles) with prices modelled (dark red) after permuting the input data. This test confirm that the input data chosen are meaningful into modeling HPI and the algorithm is not able to predict prices from noise (overfitting). The "kNN" approach has been used as for other models, tests shown in this paper.*

### 3.4.3. Sensitivity test: testing the prices using local-only parameters.

Here, I try to model the prices with nationally focused parameters. I only show in this paper the results for Switzerland and France[9], as the data for other countries were more difficult to gather. I use only three parameters here: GDP, inflation and unemployment rate.

Results of this approach are displayed for two countries (France and Switzerland) in Figure 13 and Figure 14. Those two countries were selected as 1) most of the loans are not amortized In CH and 2) the criteria of loan allocation are more demanding in Switzerland than in France (e.g. affordability parameter of 5%). I consider those models are not successful to model the prices, as neither the long-term trends nor the short term variations are recovered ( Figure 13 and Figure 14).

---

[9] Organization for Economic Co-operation and Development, Unemployment Rate: Aged 15-74: All Persons for France [LRUN74TTFRQ156N], retrieved from FRED, Federal Reserve Bank of St. Louis;

https://fred.stlouisfed.org/series/LRUN74TTFRQ156N, June 28, 2023.

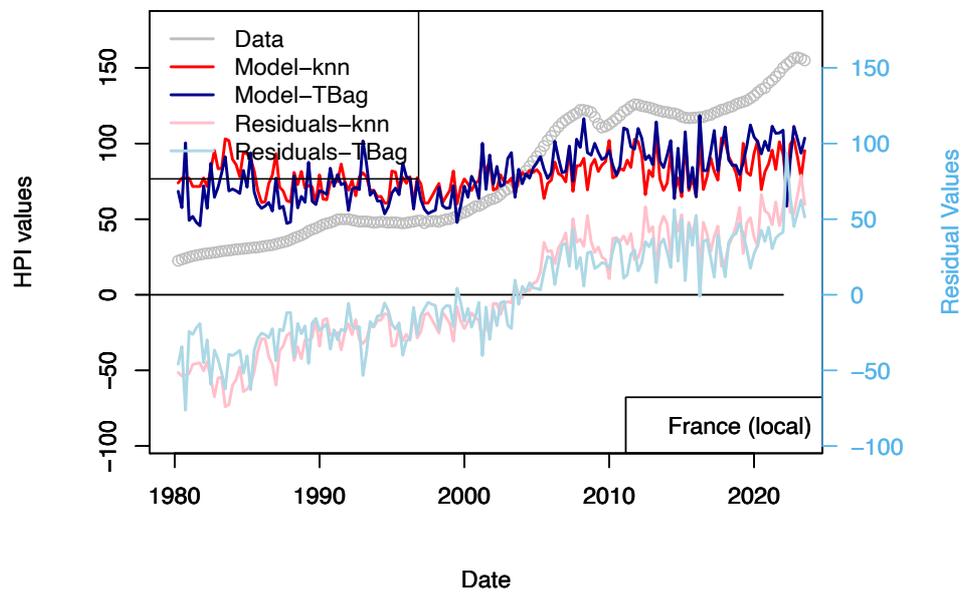

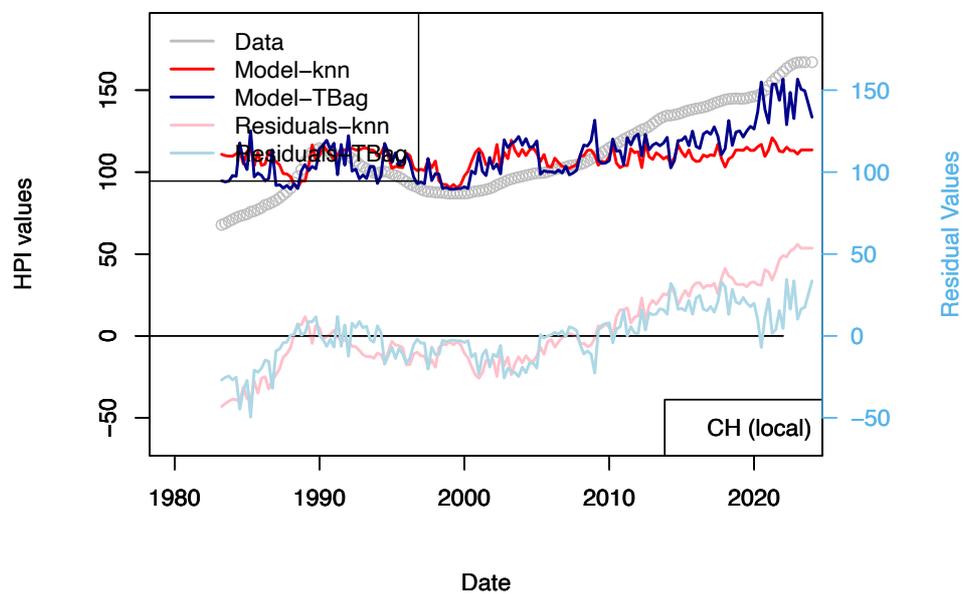

*Figure 13: "LOCAL 1yr" models for France and Switzerland. HPI are modelled using only local economic factors.*

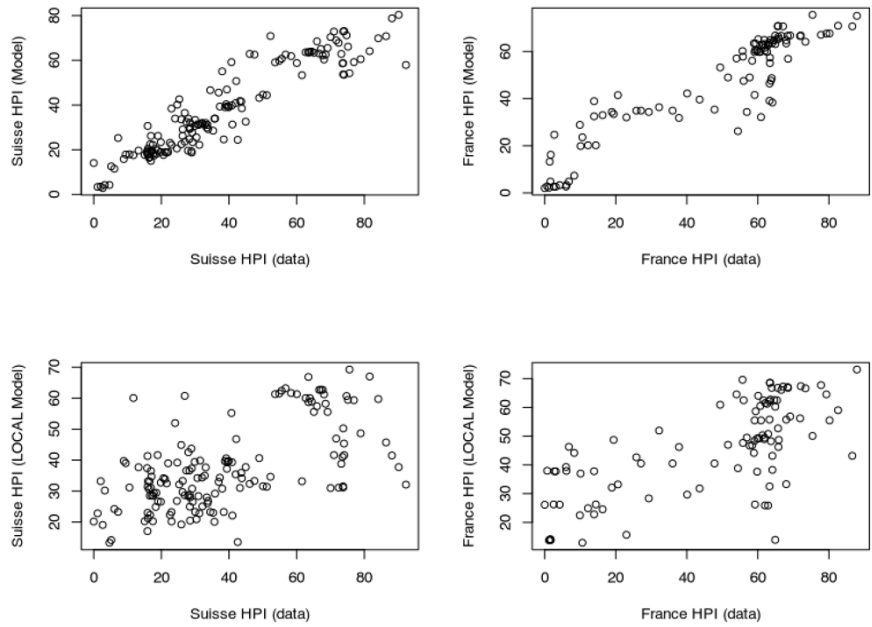

*Figure 14: Relative performance of models based on local-only dataset and models including the contribution of treasury rates ( for France and Switzerland).*

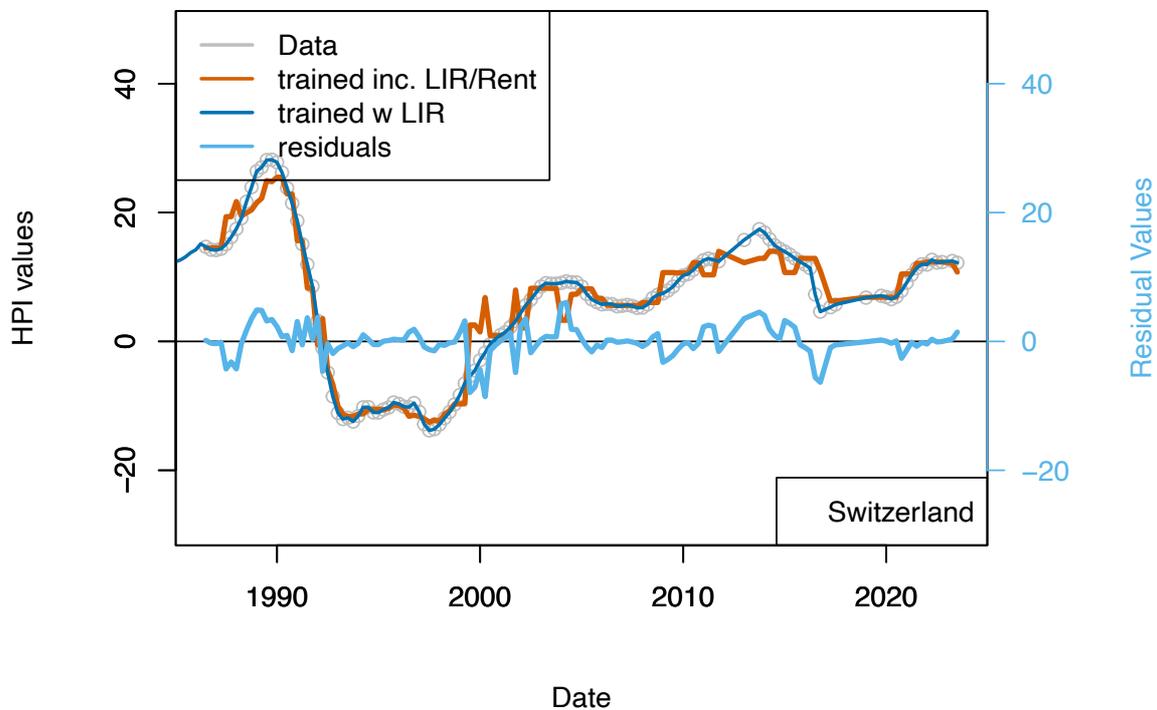

*Figure 15: "Rents 1yr" model. The model including most of the data discussed in this paper are included into the input dataset (a list of parameters is presented in Table 1). The performance of the model is compared with the model LIR (GDP, CPI, Treasury rates + central bank rates).*

### 3.4.4. Predicting price using models trained on amended datasets.

For this test, I have removed the last 4 data points from each dataset. The models are trained on this amended dataset and prices are then predicted using the complete datasets. In the current situation, the last quarter's data were strongly influenced by the action of the central banks to reduce inflation and tame the increase of wages. We nowadays see a combination of inflation and rates which was so far thought as impossible (e.g. low rates/high inflation). Now the inflation is decreasing again we reach some data space which were previously seen (e.g. in 2003). For this reason, some association of the parameters used were unprecedented.

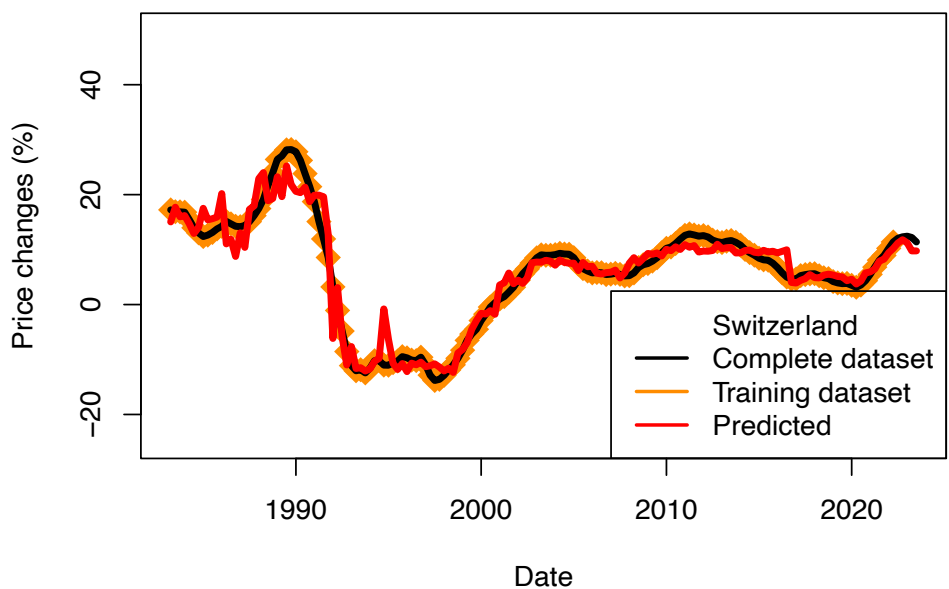

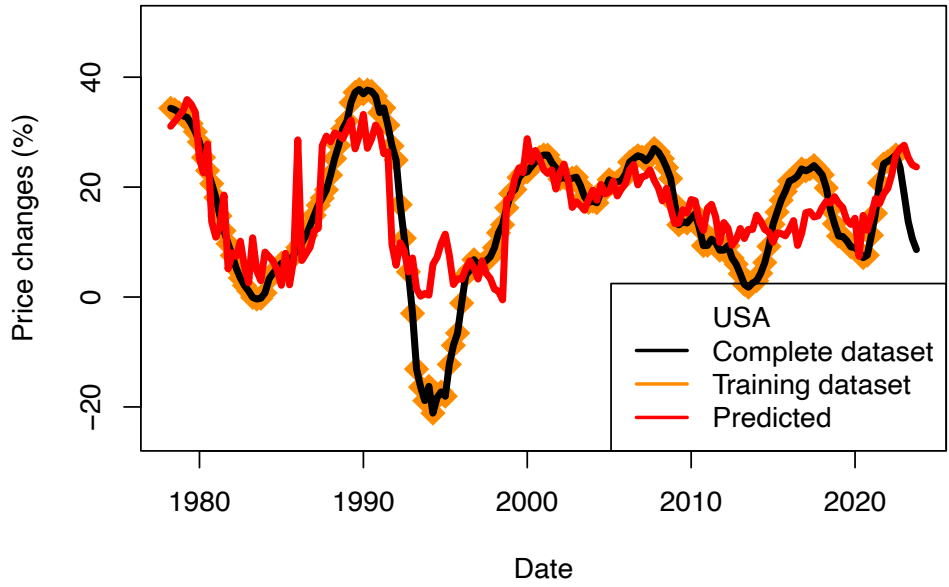

*Figure 16: Performance assessment of the price models (ECB_1y_TB_model) for 4 countries. Predicted price changes are indicated in red, complete dataset in black and training data in orange. For the 4 countries, price decrease could be modelled.*

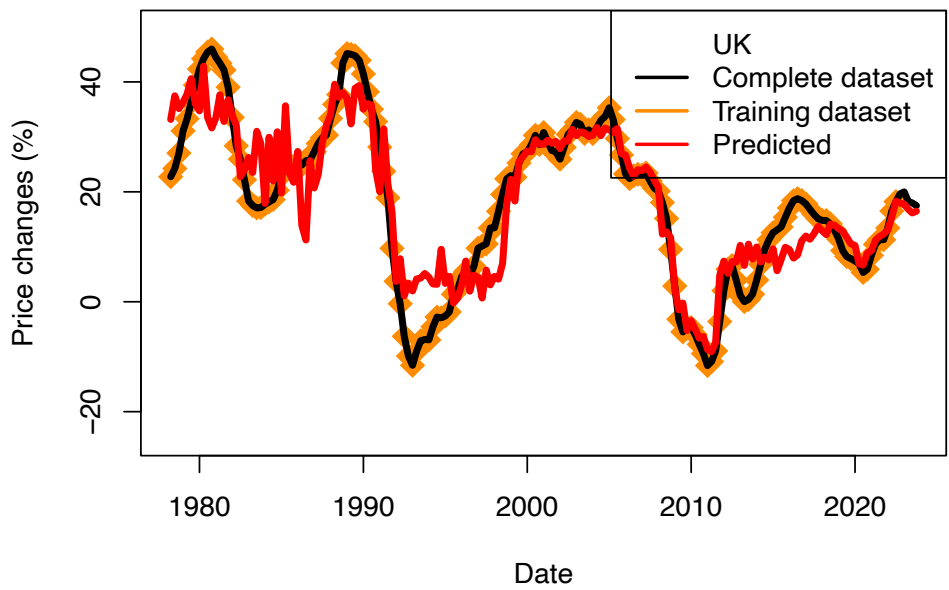

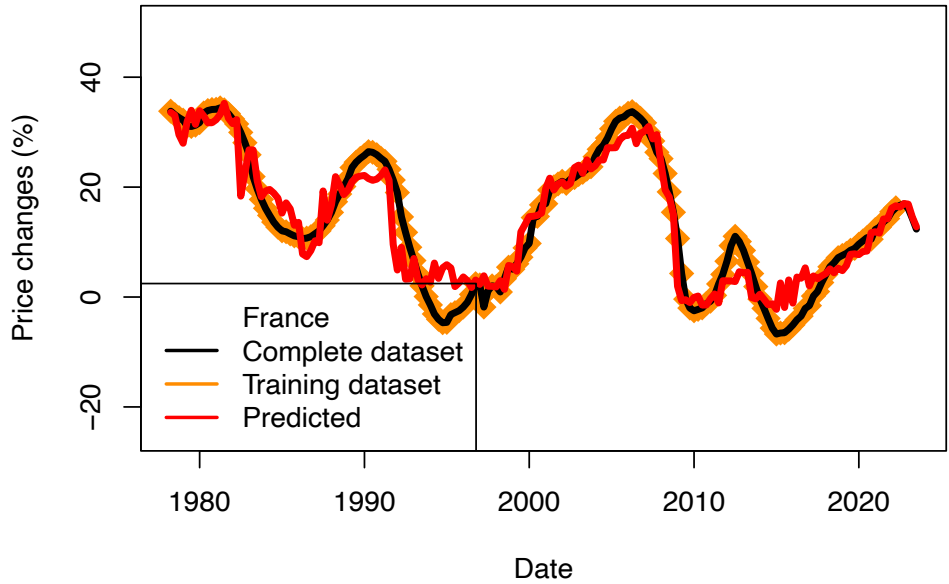

*Figure 16 (continued): Performance assessment of the price models (ECB_1y_TB_model) for 4 countries. Predicted price changes are indicated in red, complete dataset in black and s data in orange. For the 4 countries, price decrease could be modelled.*

| Country | M_COR | SD_COR | M_RMS | SD_RMS | SD_RES | M_MAE | SD_MAPE | M_CHIp | M_CHIs |
|---|---|---|---|---|---|---|---|---|---|
| USA | -0.93 | 0.34 | 10.24 | 0.66 | 4.90 | 9.32 | 0.49 | 1.00 | 0.03 |
| France | 0.84 | 0.57 | 3.28 | 0.47 | 1.42 | 3.05 | 0.25 | 0.85 | 0.15 |
| Germany | 0.72 | 0.46 | 13.17 | 0.38 | 4.99 | 12.44 | 0.65 | 1.00 | 0.03 |
| Spain | 0.27 | 0.57 | 4.43 | 0.08 | 1.29 | 4.28 | 0.74 | 1.00 | 0.03 |
| Italy | -0.62 | 0.43 | 2.90 | 0.47 | 1.76 | 2.46 | 0.61 | 0.99 | 0.04 |
| Finland | 0.94 | 0.28 | 6.59 | 0.16 | 2.62 | 6.19 | 1.54 | 1.00 | 0.03 |
| Sweden | 0.83 | 0.44 | 9.36 | 0.58 | 3.41 | 8.88 | 0.49 | 1.00 | 0.03 |
| Australia | 0.09 | 0.19 | 6.51 | 1.13 | 1.66 | 6.35 | 0.38 | 1.00 | 0.03 |
| UK | 0.57 | 0.26 | 5.98 | 0.45 | 1.10 | 5.90 | 0.51 | 1.00 | 0.03 |
| Switzerland | 0.69 | 0.15 | 3.08 | 0.32 | 1.40 | 2.83 | 0.36 | 0.83 | 0.16 |
| Netherland | 0.99 | 0.39 | 2.79 | 0.55 | 1.39 | 2.51 | 0.11 | 0.93 | 0.09 |
| Norway | 0.98 | 0.33 | 2.62 | 0.23 | 3.01 | 2.38 | 0.18 | 0.50 | 0.77 |

*Table 18: Statistics of the data fit performance of the ECB_1y_tbag models (tree-bag strategy). Here we train the model after excluding the last four data points. The predictive values are compared to the original datapoint, and statistics are computed by comparing together the 8 points. tree-bag strategy is used.*

| Country | M_COR | SD_COR | M_RMS | SD_RMS | SD_RES | M_MAE | SD_MAPE | M_CHIp | M_CHIs |
|---|---|---|---|---|---|---|---|---|---|
| USA | 0.747 | 0.000 | 10.064 | 0.203 | 4.901 | 9.516 | 0.500 | 1.000 | 0.029 |
| France | -0.829 | 0.581 | 1.873 | 0.422 | 1.416 | 1.601 | 0.112 | 0.500 | 0.771 |
| Germany | -0.723 | 0.641 | 14.887 | 0.128 | 4.986 | 14.190 | 0.681 | 1.000 | 0.029 |
| Spain | NA | 0.001 | 5.312 | 0.062 | 1.285 | 5.131 | 1.059 | 1.000 | 0.029 |
| Italy | NA | 0.001 | 3.539 | 0.001 | 1.764 | 3.069 | 0.902 | 0.750 | 0.229 |
| Finland | -0.744 | 0.279 | 7.440 | 0.069 | 2.622 | 7.017 | 1.425 | 1.000 | 0.029 |
| Sweden | 0.748 | 0.001 | 8.221 | 0.429 | 3.415 | 7.889 | 0.466 | 1.000 | 0.029 |
| Australia | NA | 0.010 | 4.160 | 1.373 | 1.657 | 3.868 | 0.191 | 0.500 | 0.771 |
| UK | -0.178 | 0.577 | 4.725 | 0.070 | 1.101 | 4.581 | 0.353 | 1.000 | 0.029 |
| Switzerland | 0.690 | 0.001 | 2.464 | 0.386 | 1.401 | 2.244 | 0.278 | 0.656 | 0.433 |
| Netherland | 0.978 | 0.001 | 4.340 | 0.830 | 1.388 | 4.296 | 0.171 | 0.852 | 0.147 |
| Norway | 0.557 | 0.000 | 3.339 | 0.078 | 3.011 | 2.536 | 0.171 | 0.500 | 0.771 |

*Table 19: Statistics of the data fit performance of the ECB_1y_kNN models (kNN strategy). Here we train the model after excluding the last four data points. The predictive values are compared to the original datapoint, and statistics are computed by comparing the 8 points together. kNN strategy is used.*

As each model is set by a random seed, I have performed 600 runs in order to build statistics shown in Tables 6 and 7. For four countries, the direction of HPI changes (negative and positive) was predicted correctly. This observation is supported by the positive values of r (see mean and standard deviation). For the US, the observed price decrease was much larger than the predicted ones (approx. 8%) for both *kNN* and *tree-bag* strategies. For both methods, the RMS are found to be lower than price dispersion between estimates or asset types. For the three other countries, the performance is better. Those results maybe contradict findings which suggested that economic fundamentals cannot fully explain the house price evolution [24, 58-60] or it simply suggest the economic data available at the time (before the most recent crises happened) were not sufficiently diverse to be tested against house price indices.

## 3.5. Predicted prices.

Now confidence is gained in predicting prices over the four last quarters, I produced four sets of MEF values to predict prices using models trained on complete datasets. The ranges of values tested represent values which could be relevant for pricing mortgages in a mid-term horizon of 4 quarters to predict values of HPI. I have used the model 6 (ECB 1yr, see Figure 11) with the tree-bag approach. Those predictions could be compared to Pillar2 scenarios simulation as required by regulators. Unlike for Pillar 2 scenarios, all possible combinations of input parameters over specified ranges (Table 20) have been considered leading to a total number of 160'000 synthetic prices. From this input dataset, I have computed prices predicted for the four countries (France, UK, USA and Switzerland). With the exception of USA, none of the models predict a decrease of price over the next 4Q (12Q metrics). This means that in those four countries, systemic margin calls are not expected to be seen in a mid-term horizon for interest only mortgages (like those available in Switzerland and USA). It should be kept in mind here that the prices predicted are bound by a maximum diminution of ECB assets of 1'600 mEUR (from a start value of 6'500mEUR). Finally, I remind you that those prices are averaged over all kinds of assets, and mean prices may be overly optimistic for some areas and some asset types depending on the local economic conditions.

| N | GDP (%) | Inflation (CPI) | ECB Asset size (M EUR) | Treasury rates (%) |
|---|---|---|---|---|
| 160'000 | -2 to 2% | -2 to 2 % | 6.5E6 +/- 1E6 | 0 to 4% |

*Table 20: Range of MEF values used to compute the predicted values (160'000 models). For each parameter, 20 values were tested over each range.*

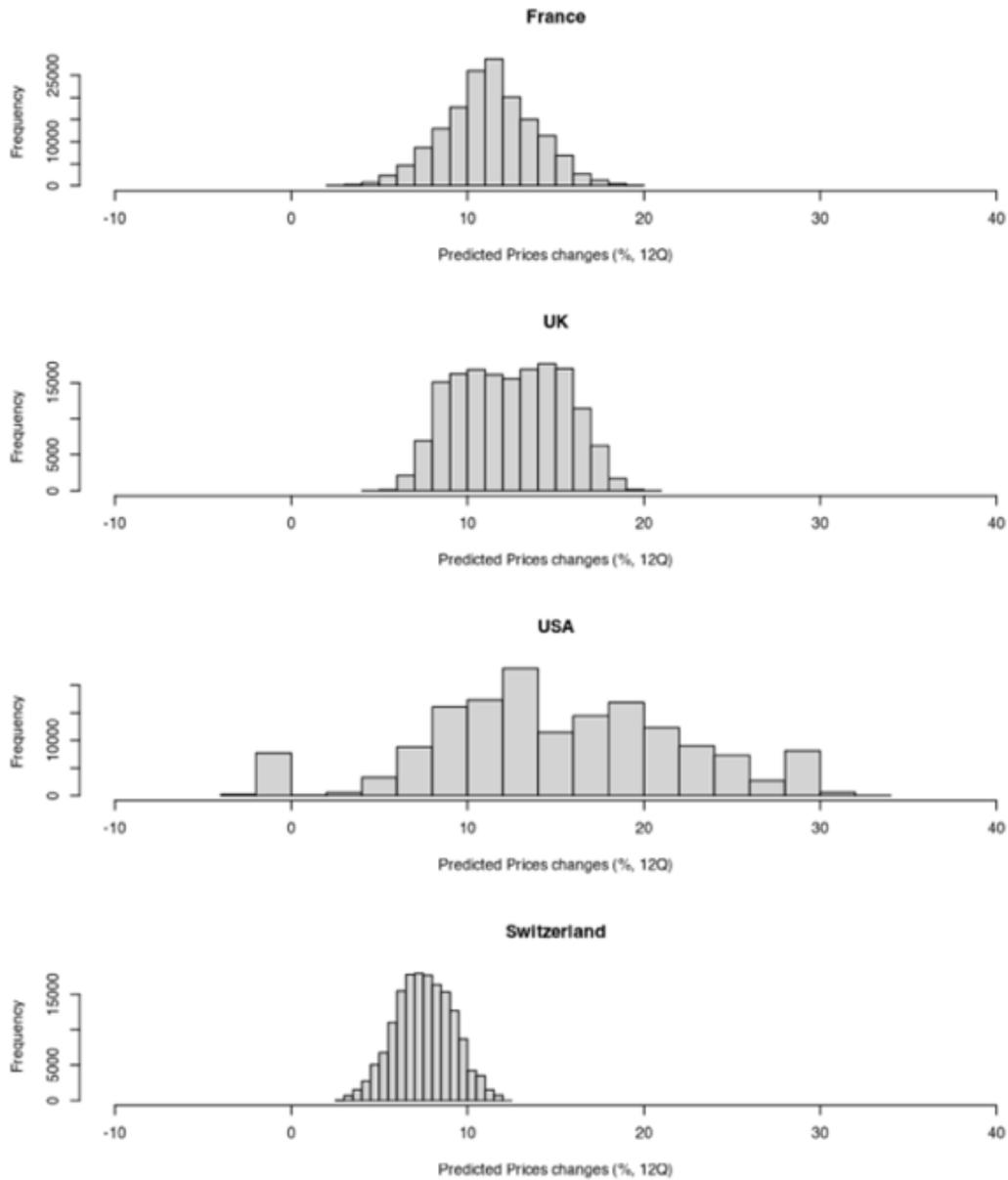

*Figure 17: Predicted price changes (over 12 quarters) for four countries. Those estimates need to be compared to the range of values of the RMS shown in the previous figure. Current values for France, UK, USA and Switzerland are 12.4%, 13.2%, 4.3%, 8.7% respectively.*

Those predictive prices do not take into account temporal development or MEF changes. Those estimates are end-of-scenario prices. The evolution of prices over n-quarters can be reconstructed using *n* prices. As the number of solutions computed is large (160'000), the risk that the future is not included in one of the scenarios is low. Such prediction allows for the computation of many scenarios, some of which may be outside economic standards models, allowing the price estimation resulting from non-conventional economic policies.

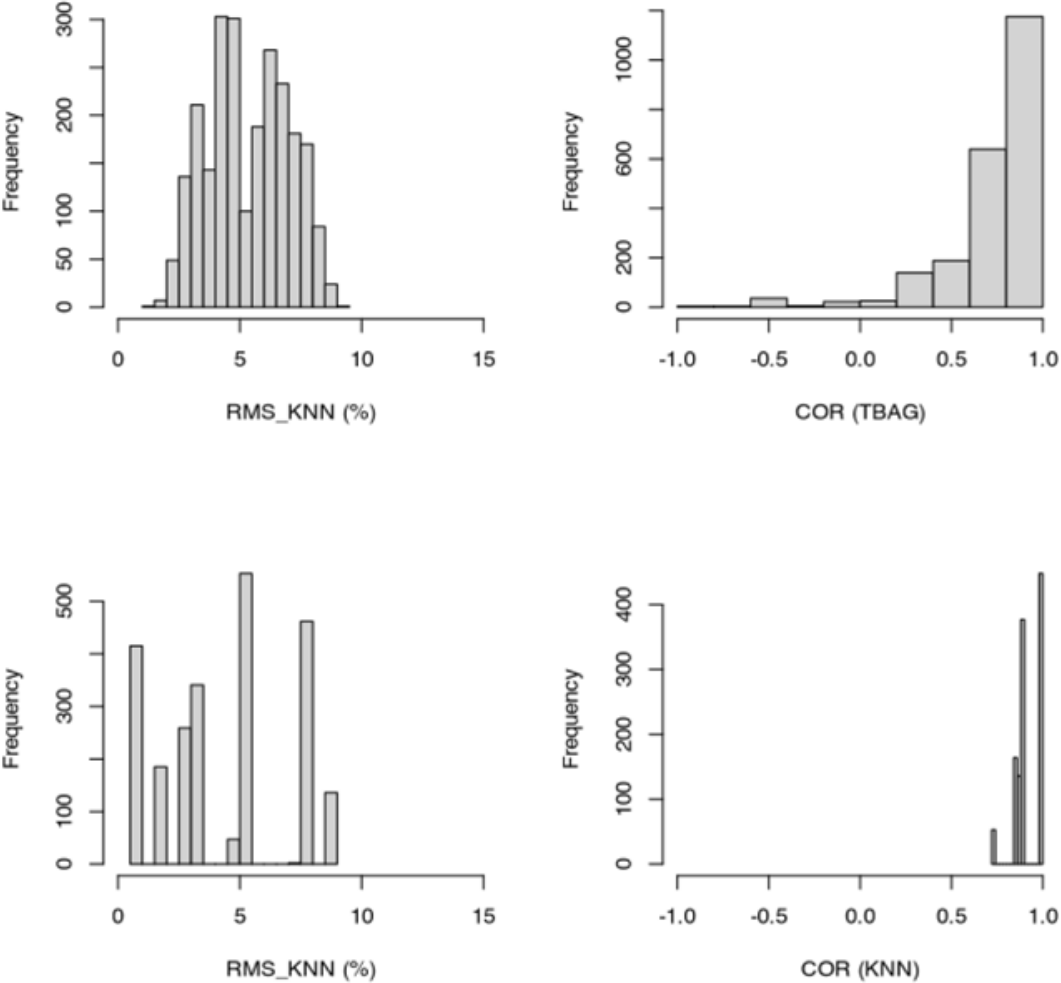

*Figure 18: Distributions of RMS and correlation parameters (COR) for the kNN and tree-bag strategies (data of the four countries are merged).*

# 4. Conclusion

A recurrent criticism of the machine learning approaches is that the results cannot be explained. I show here by completing a step-by-step modeling approach, which parameters can improve the data fit. The chosen parameters can represent all parties, enabling the transaction to be completed under certain conditions (available buyer, available seller, financing, economic support), and ultimately help understand the price structure. Even if the data quality is imperfect (e.g. GDP may have been revised many times after publication), the data quality may be improved in the future thanks to the work of independent rating agencies and the information sharing of countries with investors. Ultimately, the models are limited by the sampling rate of the dataset and the length of the period observed. Even in such a context, some models are performing well enough to predict prices over the next 4 quarters.

In summary, and as expected [61], the most critical parameter is the level of interest rates at 10 years, but, this study allows us to refine our understanding of the mechanisms in place.

- The inflation plays a role, although its contributions vary from one country to another. As the real-estate investment is often considered as an hedge to inflation, the interaction between the two variables cannot be ignored. [62-64]
- The impact of the level of employment on prices is likely stronger when the real estate market is open to many and loans are amortized (otherwise, either buyers would not have access to the loan or we would observe a credit event similar to the 2008 subprime crisis).
- When data from the ECB/FED asset purchase program are introduced, the fit with the data improves from 2015 onwards. The correlation coefficient between prices in Switzerland and ECB portfolio size is higher than 92% (computed over the last 24 quarters using the ECB nominal model; to be compared of r=46% for inflation and r=0.58 for rates; see also Table 15 and Table 16 for relative importance).

- The differentiation of the HPI allows for predicting prices for the next four quarters for at least four countries.

If I consider the models collectively, for all countries, it is possible to model prices without including the ECB portfolio size up to 2015. From 2015 onwards, the price increase and price decrease tend to correlate with the size of the ECB asset book. For some models, the importance of ECB portfolio size is equally important to the level of treasury rates. As countries outside the European community are impacted by this effect, I suggest that some large-volume investors are involved in arbitrages made between real estate, sovereign, and corporate debts investments [65-71]. The reversal of ECB policies designed to support the economy since 2015 may change the risk balance for some institutional investors and may lead to unpredictable consequences such as a risk of contagion from real estate to other financial sectors. For Switzerland, regarding the 1990 real-estate bubble, one can infer that the levels of financing and the inflation rate increase triggered the crisis. In particular, the start of the crisis coincides with the time when the ratio of CPI/TR passed 5% (downwards). The accuracy of the models presented in this study may encourage modelers involved in capital management based on economic scenarios such as ICAAP and Basel "Pillar 2 models" to use similar models to explore further the landscape of possible scenarios. In the future, and to ensure all aspects of the market are well included, I would recommend building on those models investigating more the influence of rent level (as the linkage between the buy-to-let, build-to-let, and build-to-sell markets is so far not studied enough) and by adding variables such as building rates, vacancy rates, demographic data and the age distribution of migrants. At last, ML real estate models would benefit from the comparison with models based on the capitalized income value method, where cash flows play a greater role in the pricing.

# 6. Appendix

## 6.1. Benchmarking the model against other approaches

In this section, I present benchmark tests that compare the relative performance of vector auto-regressive (VAR), linear inversion (LI) and linear models (GLM) for the same dataset (ECB_1y). This dataset has been selected as the previous analysis seemed to be the most promising of all. I remind here that the goal of those tests is not to find the most relevant parameters as in the rest of the text but only to compare the performance of all approaches with the dataset ECB_1yr.

*Vector auto-regressive model (VAR)*

First I test the capabilities of the VAR model to predict values of the dataset ECB_1y. In order to predict the next values I did not use the default function options which extrapolate input parameters but specified explicitly the input data of the observed dataset as input parameter of the R::predict function [72]. I complete the prediction tests for various period starting with a test using 50 data dates to predict the next 111 values of HPI rates. I have made all other test for each 10 additional point (50, …110, 120, 130…) until there is only 4 quarters left to predict. For all models, I have searched for both a constant and trend (i.e. options type="both" and lag value p = 2 have been used). All models are shown in Figure 19.

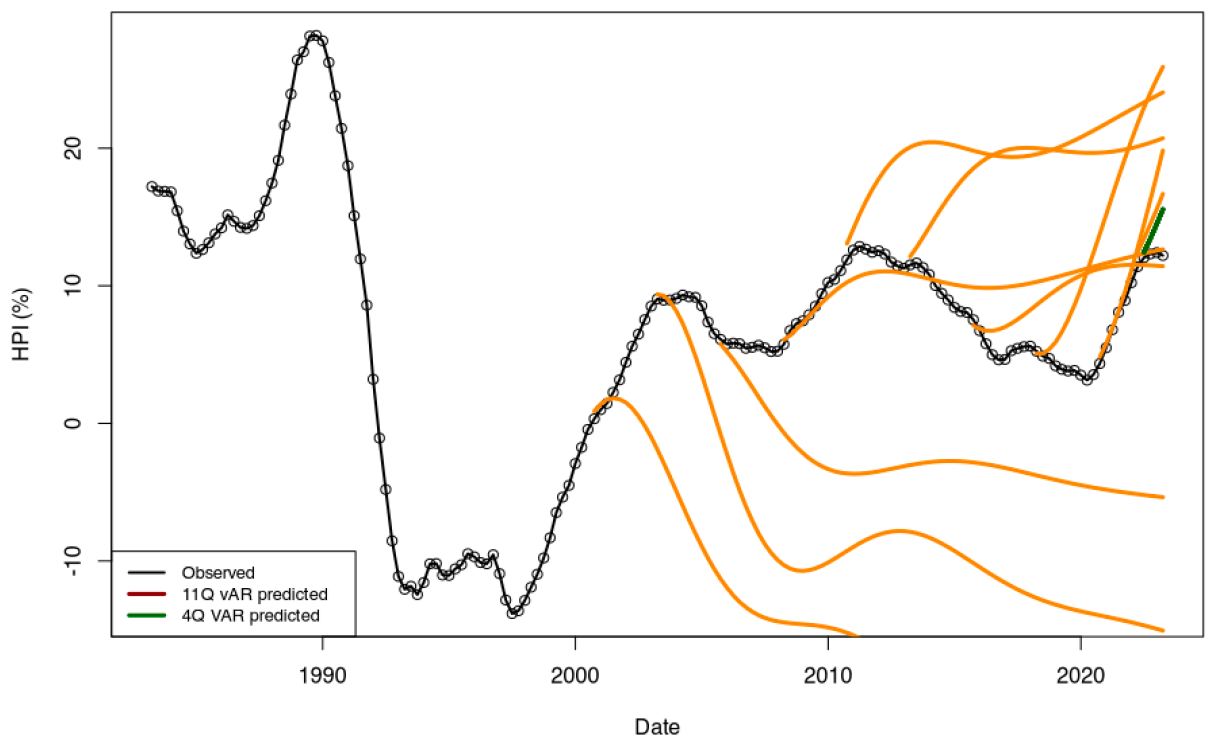

*Figure 19: Time-series recovered using R::caret package and the ECB_1y dataset. From the 161 points available, the model determined are used to predict the rest of the time-serie. Predictions (dark orange curves) were made using 100, 110, 120, … data points. The performance of the first points after the last training point, is not satisfactory, even for the model with most data (157 data points used, 4 points predicted, in dark green color).*

From visual inspection, none of these tests is performant. Figure 19 shows as expected the VAR approach is only functional when the input parameters are stationary independently and together. As this is not the case (CPI and financing rates are not always well connected, GDP decrease is not linked with HPI rate, etc...) the HPI rates cannot be modelled efficiently. Time-series shown suggest that in this situation the VAR is only able predict the extrapolation of the most recent context.

*Linear inversion (LI)*

I have performed a linear inversion of the house prices [73] for Switzerland (ECB_1y dataset):

$$\begin{bmatrix} HPI_1 \\ ... \\ HPI_n \end{bmatrix} = A \begin{bmatrix} ECB_1 & \cdots & CPI_1 \\ ... & \cdots & ... \\ ECB_n & \cdots & CPI_n \end{bmatrix}$$

here the matrix $A$ stores the coefficients of the best fitting model and the right-hand matrix stores the data used to model index values (HPI). In this situation, each data has the same weight.

| ECB portfolio size | CPI | GDP | Treasury rate (10y) |
|---|---|---|---|
| 1.02E-6 | 1.15 | 1.42 | 0.65 |

*Table 21: Parameters recovered from the linear inversion of the data.*

The parameters are presented in Figure 20. The poor fit between computed values and original values can be explained by the fact that a same HPI rates can be explained by different context and set of values at two different dates.

From the data fit shown in Figure 20, it may appear not necessary to compute the predicted values from the model obtained from this method (as no model without a good fit could be expected to provide reliable prediction values). However, estimates of prices for this method are shown in Table 23.

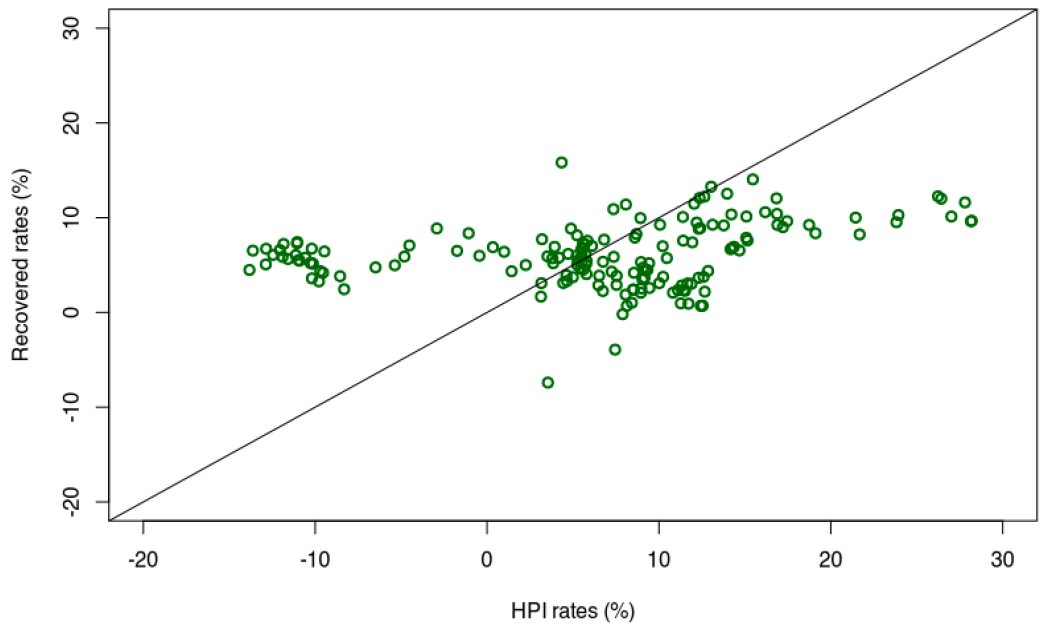

*Figure 20: Values recovered from the linear combination of input values associated with the coefficients presented in Table 21. The 1:1 relationship is indicated with a straight line.*

*Linear model (GLM)*

At last, I completed a linear model using the R::glm function and the ECB_1y dataset. In order to show the dispersion of the solution and apprehend error bars, I have run the model 600 times using a random normally distributed disturbances within +/- 10% range for each input parameters. Recovered time-series are shown in Figure 21 in grey. The performance of the models is overall poor and as for the linear inversion, it could be explained by the fact a single price has no unique explanation. This hypothesis is supported by the fact that most recent estimates are more scattered than older ones. As noted in the main text, the financial context changed first in 2015 (ECB intervention) and then again in 2021 (shrinking of the ECB portfolio). As those new contexts were less documented they have a lower weights in the solutions obtained from the complete dataset.

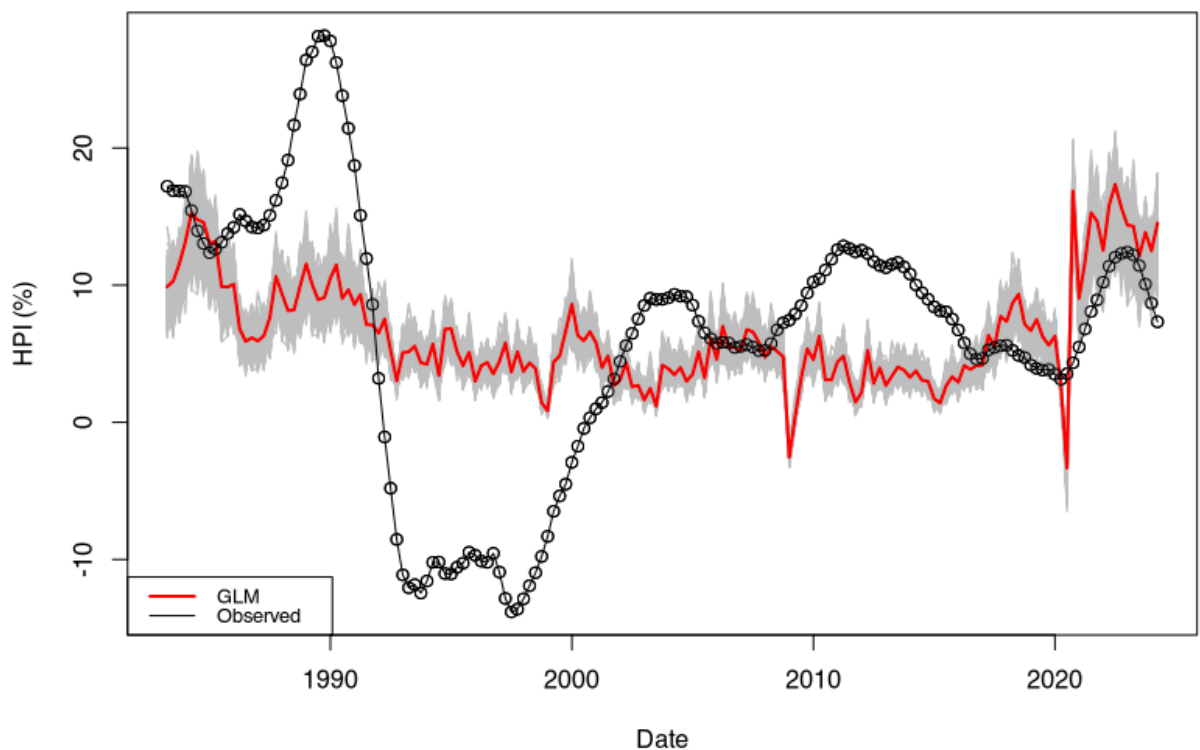

*Figure 21: Time-series recovered from the R::glm function. The model has been designed on the first 161 values of the dataset ECB_1y. The undisturbed parameter is shown using a red line. The time-series obtained from the "disturbed" datasets are shown in grey. The original HPI time-series is plotted using red circles.*

| Parameter | intercept | ECB | CPI | GDP | TR_10y |
|---|---|---|---|---|---|
| Value | -3.21 +/- 0.24 | 1.44 E-6 +/- 2.95E-8 | 0.88 +/- 0.11 | 1.37 +/- 0.11 | 1.10 +/- 0.04 |

*Table 22: Parameters constrained using the R::glm approach. Those values were disturbed 600 times using 10% amplitudes ranges.*

In summary, I computed the price predictions for the last fours quarters of the dataset using the various methods. Results shown in Table 23 suggest tested alternative methods are not able to predict a decrease of prices from 2023-Q2 onwards.

| Method | 2023-Q2 | 2023-Q3 | 2023-Q4 | 2024-Q1 |
|---|---|---|---|---|
| VAR (p=2) | 11.91 +/- 1.08 | 11.57 +/- 2.29 | 11.25 +/- 3.63 | 10.93 +/- 5.05 |
| Linear Inversion | 11.36 | 13.07 | 11.80 | 13.83 |
| GLM | 12.12 | 13.85 | 12.51 | 14.52 |
| ML kNN | 9.68 | 7.38 | 6.47 | 6.47 |

| Data Observed | 11.42 | 10.07 | 8.69 | 7.32 |

*Table 23: Recovered values for all models considered in this section, plus the values obtained from the prediction based on the ML kNN model. The ML model is the only model which is able to predict a clear decrease of prices over the next 4 quarters (starting 2023-Q2.*